\documentclass[longauth]{aa}

\usepackage{graphicx}
\usepackage{txfonts}
\usepackage{multirow}
\usepackage{multicol}
\usepackage{lscape}
\usepackage{booktabs}
\usepackage{threeparttablex}
\usepackage{wasysym}
\usepackage{xcolor}
\usepackage{array}
\usepackage{arydshln}
\usepackage{siunitx}
\usepackage[bookmarks=false,colorlinks=true,linkcolor=blue,citecolor=blue,filecolor=blue,urlcolor=blue]{hyperref}

\definecolor{kd}{rgb}{0.0,0.6,0.0}
\definecolor{kdskiptext}{rgb}{0.7,0.7,0.7}

\definecolor{sf}{rgb}{0.6,0.0,1.0}

\definecolor{gl}{rgb}{1.0,0.2,0.2}

\definecolor{js}{rgb}{0.5,0.,0.6}

\begin{document} 

\title{The SRG/eROSITA All-Sky Survey DR2}
\subtitle{Cumulative X-ray catalogues from the first three surveys and multi-wavelength counterparts in the western Galactic hemisphere}

\author{M.~E.~Ramos-Ceja\inst{1}\thanks{e-mail: mramos@mpe.mpg.de}, 
G.~Lamer\inst{2},
M.~Salvato\inst{1,3}, 
A.~Merloni\inst{1}, 
J.~S.~Sanders\inst{1},
A.~Georgakakis\inst{4},
T.~Liu\inst{5,6},
E.~Bulbul\inst{1}, 
J.~Buchner\inst{1}, 
K.~Dennerl\inst{1}, 
M.~J.~Freyberg\inst{1}, 
S.~Friedrich\inst{1}, 
I.~Kreykenbohm\inst{7}, 
C.~Maitra\inst{8,1}, 
K.~Nandra\inst{1}, 
P.~Predehl\inst{1}, 
T.~H.~Reiprich\inst{12}, 
J.~Robrade\inst{9}\thanks{This paper is dedicated to the memory of our dear colleague Jan Robrade, who recently passed away. His substantial contributions to the eROSITA mission and his warmth as a colleague will be remembered with gratitude.}, 
A.~Schwope\inst{2}, 
R.~Shirley\inst{1}, 
B.~Stelzer\inst{10}, 
I.~Stewart\inst{1}, 
R.~Seppi\inst{11},
H.~Starck\inst{1}, 
D.~Tub{\'i}n-Arenas\inst{2}, 
I.~Traulsen\inst{2},
E.~Artis\inst{1},
C.~Aydar\inst{1},
P.~Baldini\inst{1},
F.~Balzer\inst{1},
W.~Becker\inst{1,13},
M.~M.~Bennedik\inst{10},
W.~Bornemann\inst{1},
M.~Br{\"u}ggen\inst{9},
J.~Brink\inst{2,14},
M.~Brusa\inst{15,16},
V.~Burwitz\inst{1},
M.~Canal~i~Saguer\inst{7},
N.~Clerc\inst{17},
J.~Comparat\inst{18},
M.~Coriat\inst{17},
J.~V.~Corr\^{e}a-Rodrigues\inst{7,19},
S.~Czesla\inst{9,20},
L.~Dauner\inst{7},
J.~Dietl\inst{12},
Z.~Ding\inst{1},
L.~Ducci\inst{10},
T.~Dwelly\inst{1},
L.~Fiorino\inst{1},
S.~Freund\inst{1},
P.~Friedrich\inst{1},
R.~Gaida\inst{1},
E.~Gatuzz\inst{1},
S.~T.~Guida\inst{1},
S.~H{\"a}mmerich\inst{7},
F.~Haberl\inst{1},
G.~Hartner\inst{1},
S.~Hern\'andez-D\'iaz\inst{10},
Z.~Igo\inst{1},
N.~Ilic\inst{2},
D.~M.~Kaltenbrunner\inst{1},
A.~Khokhriakova\inst{1},
W.~Kink\inst{1},
C.~Kirsch\inst{7},
M.~Kluge\inst{1},
S.~Krippendorf\inst{21},
M.~Krumpe\inst{2},
S.~Kulkarni\inst{1},
J.~Kurpas\inst{2},
E.~Kyritsis\inst{1},
R.~Laktionov\inst{7},
A.~Liu\inst{22,23},
M.~Lorenz\inst{7},
N.~Malavasi\inst{1},
M.~G.~F.~Mayer\inst{7},
N.~Meidinger\inst{1},
T.~Mistele\inst{1},
S.~M{\"u}ller\inst{1},
D.~Muñoz-Giraldo\inst{10},
N.~T.~Nguyen-Dang\inst{10,25},
Q.~Ni\inst{1},
S.~Ok\inst{2,24},
N.~Ota\inst{12,25},
G.~P{\"u}hlhofer\inst{10},
F.~Pacaud\inst{12},
A.~Pandya\inst{10},
E.~Perinati\inst{10},
C.~Pommranz\inst{10},
G.~Ponti\inst{26,1,27},
K.~Poppenhaeger\inst{2,14},
K.~G.~Pradeep\inst{2,14},
A.~Rau\inst{1},
W.~Roster\inst{1},
S.~Rukdee\inst{1},
S.~Saeedi\inst{7},
A.~Santangelo\inst{10},
M.~Sasaki\inst{7},
S.~Sheth\inst{2,14},
S.~Shreeram\inst{1},
M.~Sommer\inst{1},
A.~Srivastava\inst{28,12},
V.~Suleimanov\inst{10},
J.~Tr{\"u}mper\inst{1},
N.~Vasilas\inst{1},
A.~Veronica\inst{12},
N.~Webb\inst{17},
P.~Weber\inst{7},
J.~Wilms\inst{7},
M.~C.~H.~Yeung\inst{1},
F.~Zangrandi\inst{7},
S.~Zelmer\inst{1},
X.~Zhang\inst{1},
Y.~Zhang\inst{29},
X.~Zheng\inst{1}
}
\institute{
Max-Planck-Institut f\"ur extraterrestrische Physik, Gie{\ss}enbachstra{\ss}e, D-85748 Garching, Germany
\and 
Leibniz Institut f\"ur Astrophysik Potsdam, An der Sternwarte 16, D-14482 Potsdam, Germany
\and
Exzellenzcluster ORIGINS, Boltzmannstr. 2, D-85748 Garching, Germany
\and
Institute for Astronomy and Astrophysics, National Observatory of Athens, V. Paulou and I. Metaxa, 11532 Athens, Greece
\and
Department of Astronomy, University of Science and Technology of China, Hefei 230026, People’s Republic of China
\and
School of Astronomy and Space Science, University of Science and Technology of China, Hefei 230026, People’s Republic of China
\and 
Dr. Karl Remeis-Sternwarte and Erlangen Centre for Astroparticle Physics, Friedrich-Alexander Universität Erlangen-Nürnberg, Sternwartstra{\ss}e 7, 96049 Bamberg, Germany
\and
Inter University Centre for Astronomy \& Astrophysics, Ganeshkhind, Pune 411007, India
\and
Hamburger Sternwarte, University of Hamburg, Gojenbergsweg 112, 21029 Hamburg, Germany
\and 
Institut f\"ur Astronomie und Astrophysik, Universit\"at T\"ubingen, Sand 1, D-72076 T\"ubingen, Germany
\and
Department of Astronomy, University of Geneva, Ch. d’Ecogia 16, CH-1290 Versoix, Switzerland
\and
Argelander Institute for Astronomy, University of Bonn, Auf dem H{\"u}gel 71, 53121 Bonn, Germany
\and
Max-Planck-Institut f{\"u}r Radioastronomie, Auf dem H{\"u}gel 69,53121 Bonn, Germany
\and
Institute for Physics and Astronomy, University of Potsdam, Karl-Liebknecht-Str. 24/25, 14476 Potsdam, Germany
\and
Dipartimento di Fisica e Astronomia "Augusto Righi", Universit{\`a} di Bologna, via Gobetti 93/2, 40129 Bologna, Italy
\and
INAF - Osservatorio di Astrofisica e Scienza dello Spazio di Bologna, via Gobetti 93/3, 40129 Bologna, Italy
\and
IRAP, Universit{'e} de Toulouse, CNRS, UPS, CNES, Toulouse, France
\and
Univ. Grenoble Alpes, CNRS, Grenoble INP, LPSC-IN2P3, 53, Avenue des Martyrs, 38000, Grenoble, France
\and
Universidade de S\~{a}o Paulo, Instituto de Astronomia, Geof\'{i}sica e Ciências Atmosf\'{e}ricas, Rua do Mat\~{a}o 1226, S\~{a}o Paulo, SP, 05508-090, Brazil
\and
Th\"uringer Landessternwarte Tautenburg, Sternwarte 5, 07778 Tautenburg
\and
DAMTP and Department of Physics, University of Cambridge, Wilberforce Road, Cambridge CB3 0WA, United Kingdom
\and
School of Physics and Astronomy, Beijing Normal University, Beijing 100875, China
\and
Institute for Frontiers in Astronomy and Astrophysics, Beijing Normal University, Beijing 102206, China
\and
Department of Astronomy \& Space Sciences, Faculty of Science, University of Ege, 35100 Bornova, Izmir, T\"urkiye
\and
Department of Physics, Nara Women’s University, Kitauoyanishi-machi, Nara, Nara 630-8506, Japan
\and
Osservatorio Astronomico di Brera, (INAF), Via E. Bianchi 46, Merate, 23807, Italy
\and
Como Lake Center for Astrophysics (CLAP), DiSAT, Università degli Studi dell’Insubria, via Valleggio 11, 22100 Como, Italy
\and
Dipartimento di Fisica, Università degli studi di Roma ‘Tor Vergata’, Via della Ricerca Scientifica, 1, 00133 Roma, Italy
\and
Max Planck Institute for Astrophysics, Karl-Schwarzschild-Str. 1, D-85741 Garching, Germany
}

\titlerunning{The eRASS:3 catalogues and their multi-wavelength counterparts}
\authorrunning{Ramos-Ceja, M. E., et al.}

\abstract{
The eROSITA telescope array on board the Spektrum-Roentgen-Gamma (SRG) mission began its all-sky survey program in December 2019, scanning the sky at an approximately six-month cadence. Here, we present new catalogues of point-like and extended X-ray sources derived from the first three completed eROSITA all-sky surveys (eRASS:3), covering the western Galactic hemisphere, which Germany's eROSITA consortium holds proprietary rights for. We describe the observational strategy, data processing, and analysis pipelines. We also characterise the resulting X-ray source populations. The eRASS:3 main catalogue contains nearly two million sources, including \num{1911744} point-like and \num{63796} extended sources, detected in the $0.2-2.3$~keV energy band, eROSITA’s most sensitive energy window. This volume has doubled the X-ray source content of eRASS1 and provides a comprehensive census of X-ray-emitting objects across diverse astrophysical classes. We also released a secondary hard catalogue of nearly \num{15000} sources detected in the harder $2.3$–$5.0$~keV energy band. In addition, we released six catalogues, three for the main sample and three for the hard sample, in which we identify and classify the optical and infrared counterparts of eRASS:3 point-like sources. This approach delivers a homogeneous, high-quality identification of mostly extragalactic X-ray emitters; for example, we estimate that $\sim88\%$ of the $1.4$ million counterparts identified within the footprints of the Legacy Survey Imaging for DESI are extragalactic sources. This work has enabled the generation of samples optimised for completeness and purity, while expanding the discovery space for rare populations. The second data release (DR2) of the SRG/eROSITA all-sky survey is a catalogue-only release, comprising the catalogues presented in this work together with an updated version of the eROSITA upper flux limit server.
}


\maketitle

%
%

%
\section{Introduction}
\label{sec:intro}
The extended ROentgen Survey with an Imaging Telescope Array \citep[eROSITA,][]{Predehl2021}, aboard the Russian–German Spektrum-Roentgen-Gamma (SRG) mission \citep{Sunyaev2021}, has carried out the most sensitive imaging all-sky survey in the soft X-ray band since the ROSAT mission in 1990 \citep{Truemper1982}, reaching a sensitivity in the $0.2-2.3$~keV energy band approximately 25 times greater than that achieved by ROSAT. Launched in July 2019 and placed in a large halo orbit around the second Lagrange point (L2), eROSITA has performed four complete all-sky surveys over two years. Its unique combination of large grasp, good angular resolution, broad spectral coverage ($0.2-8.0$~keV), and low instrumental background has enabled transformative studies across a broad range of astrophysical topics, from the growth of supermassive black holes and the formation of large-scale structure to stellar activity, the hot phase of the interstellar medium and of galaxies, and cosmology.

In brief, eROSITA comprises seven co-aligned Wolter-I mirror modules, each paired with a CCD detector. This modular design provides redundancy and has been key to the observatory's high observational efficiency during the active phases of its science operations. Equipped with a wide field of view approximately $1^\circ$ in diameter, the eROSITA telescope achieves an on-axis angular resolution of $\sim18\arcsec$ in terms of the half energy width (HEW), with an average HEW across the field of $\sim30\arcsec$ in the $0.2-2.3$~keV energy band \citep{Merloni2024}. Its effective area peaks at $\sim1365$~cm$^{2}$ at 1~keV and its time resolution is $50$~ms. We refer to the following publications and references therein for a detailed technical description of the eROSITA instrument: \cite{Predehl2021,Sunyaev2021,Coutinho2022,Merloni2024}.

Rights for the eROSITA survey data are divided equally between two consortia, with observations in the eastern Galactic hemisphere assigned to the Russian consortium and those in the western Galactic hemisphere assigned to the German collaboration \citep{Merloni2024}. This paper describes the source catalogues derived from the first three all-sky survey data obtained in the western Galactic hemisphere.

\begin{table*}
    \centering
    \caption{Timeline of the main eROSITA operations and major events during eRASS2 and eRASS3 \citep[see][for details on eRASS1]{Merloni2024}.}
    \renewcommand{\arraystretch}{1.2}
    \label{tab:erass23_operations}
    \begin{tabular}{>{\centering\arraybackslash}p{3cm}|p{13cm}}
        \hline
        \hline
        Date & Description \\
        \hline
         11.06.2020 & Start of eRASS2. \\
         16.06.2020 & Routine orbit correction (for station keeping). \\
         18.06.2020 & Observation of 1E\,0102$-$72 for soft energy calibration. \\
         05.08.2020 & Routine orbit correction (for station keeping). \\
         06.08.2020 & Observation of 1RXS J160518.8+324907 for contamination monitoring. \\
         05.10.2020 & First major orbit correction and controlled ITC reset. \\
         07.10.2020 & Observation of 1RXS J185635.1$-$375433 for contamination monitoring. \\
         10-16.11.2020 & In-flight calibration measurements (one camera per day). \\
         23.11.2020 & Second major orbit correction. eROSITA to safe mode. Cameras switched off. \\
         25.11.2020 & Observation of 1RXS J214303.7+0655419 for contamination monitoring. \\
         26.11.2020 & Major ITC anomaly. \\
         15.12.2020 & End of eRASS2. \\
        \hline
         15.12.2020 & Start of eRASS3. \\
         29.01.2021 & Micrometeoroid impact on TM7. \\
         01-11.02.2021 & In-flight calibration measurements (one camera per day). \\
         23.02.2021 & Micrometeoroid impact on TM4. \\
         28.02.2021 & Third major orbit correction. eROSITA to safe mode. Cameras switched off. \\
         07.03.2021 & Observation of 1RXS J160518.8+324907 for contamination monitoring. \\
         10.03.2021 & Micrometeoroid impact on TM5. \\
         15.04-20.05.2021 & ESTRACK\tablefootmark{a} support due to low visibility from Russian antennas. \\
         17.05.2021 & Micrometeoroid impact on TM2. \\
         23.05.2021 & Fourth major orbit correction. eROSITA to safe mode. Cameras switched off. \\
         25.05.2021 & Observation of 1RXS J214303.7+0655419 for contamination monitoring. \\
         16.06.2021 & End of eRASS3. \\
        \hline
    \end{tabular}
\tablefoot{
\tablefoottext{a}{Due to a low spacecraft visibility window during the spring period of 2021 \citep{Sunyaev2021}, two deep-space antennas of the ESA
tracking stations network (ESTRACK) were also used on 12 occasions for the science data downlink of about $5.7$~GB.}
}
\end{table*}

Throughout its main phase of scientific operations, eROSITA has adopted a scanning strategy that allows it to cover the whole sky every six months. \cite{Merloni2024} presented a comprehensive source catalogue and scientific overview from the first of these passes, known as the first eROSITA All-Sky Survey (eRASS1), which marked a pivotal step in achieving the mission’s science goals. eRASS1 achieved a median sensitivity (at around 50\% completeness) of $\sim 5\times10^{-14}$~erg~cm$^{-2}$~s$^{-1}$ in the $0.5-2.0$~keV energy band across the sky, detecting nearly one million X-ray sources over half of the sky — the vast majority of which are active galactic nuclei \citep[AGNs,][]{Salvato2025}, along with a significant population of stars, clusters of galaxies, and compact objects \citep{Seppi2022, Bulbul2024, Merloni2024}. Moreover, \cite{Merloni2024} detailed the survey strategy, calibration procedures, source detection algorithms, and catalogue construction. Subsequent follow-up works have also showcased the initial scientific potential of eRASS1, including variability studies, AGN population statistics \citep[e.g.][]{Bogensberger2024, Wolf2024, Waddell2026}, and the detection of galaxy clusters across a wide redshift range \citep{Bulbul2024, Kluge2024}. The catalogues presented in \cite{Merloni2024} were part of the First Data Release\footnote{\url{https://erosita.mpe.mpg.de/dr1}} (DR1) of the SRG/eROSITA all-sky survey, which also included calibrated event files, source products (light curves and spectra), and all-sky maps. As the first in a planned series of eight surveys, eRASS1 laid the groundwork for deeper and more refined catalogues in subsequent data releases.

The first three all-sky surveys, known as eRASS1, eRASS2, and eRASS3\footnote{eRASS($n$) means ($n$th) eROSITA All-Sky Survey.}, progressively increase the resulting catalogues' exposure time, depth, and completeness, since each subsequent survey benefits from increased exposure, improved calibration, and refined data processing algorithms. The cumulative eRASS:3\footnote{eRASS:$n$ means cumulative dataset from the first consecutive $n$ eROSITA All-Sky Surveys.} survey spans observations from December 2019 to June 2021. eRASS:3 offers a unique combination of sensitivity, sky coverage, and temporal information. As we demonstrate here, in the soft energy band ($0.5-2.0$~keV), the average point-source sensitivity reaches $\sim 2.7\times10^{-14}$~erg~cm$^{-2}$~s$^{-1}$, representing a substantial improvement over eRASS1 (see Sect.~\ref{sec:flux_limit}). This depth has enabled the detection of almost two million X-ray sources, including AGNs out to high redshift, tens of thousands of galaxy clusters out to $z>1$, and a rich population of Galactic stars, compact objects, as well as diffuse emission structures.

Compared to eRASS1, eRASS:3 offers enhanced capabilities for characterising source properties. The increased photon statistics from bright sources enable better spectral fitting, measurements of the source extent, and studies of short- to medium-timescale variability. For example, improvements are significant for AGN science, where flux variability and spectral shape are key diagnostics of accretion physics and obscuration, and for galaxy clusters, whose detection and characterisation benefit from refined estimates of temperature, luminosity, and morphology. In the Galactic plane, the survey resolves complex regions of diffuse emission and identifies numerous compact sources, including stars, X-ray binaries, and cataclysmic variables.

An additional strength of eRASS:3 lies in its synergy with multi-wavelength datasets. With improved astrometric accuracy ($< 1\arcsec$ for bright sources) and deeper flux limits, the cumulative eRASS:3 survey enables a more reliable cross-matching with deep optical (e.g. SDSS-V, DES, Pan-STARRS, HSC), infrared (e.g. WISE, VISTA), and radio (e.g. LOFAR, ASKAP, VLASS) catalogues. This is crucial for building statistically complete samples of different source populations and constraining their physical properties through broadband spectral energy distributions, photometric redshifts, and environmental metrics. Moreover, a substantial fraction of the eRASS:3 source counterparts have been observed spectroscopically by SDSS-V \citep{Kollmeier2026} as part of the Black Hole Mapper program. In its 20th data release (Griffith et al., in prep.), approximately 200\,000 eRASS:3 X-ray sources are expected to have high-quality optical spectra and robust redshift determinations (Merloni et al., in prep.; Roster et al., in prep.). This unprecedented level of coordinated X-ray and spectroscopic information considerably enhances the legacy value and broad community impact of the eRASS:3 data.

This paper presents the eRASS:3 source catalogues, along with the accompanying catalogues of optical counterparts and classification for the X-ray point sources. The cumulative eRASS:3 survey dataset forms the basis of this work, whereas eROSITA-DE Data Release 2 (DR2) comprises the publicly released source catalogues derived from it. In Sect.~\ref{sec:operations}, we describe the observing operations that took place during the eRASS:3 period. Section~\ref{sec:cal_processing} summarises the main aspects of data processing and its calibration. Detailed information on the software system and calibration has already been presented in \cite{Brunner2022} and \cite{Merloni2024}. We refer to those works for more information. In Sect.~\ref{sec:erass3_catalogs}, we present the eRASS:3 catalogues produced by our standard processing pipeline. Section~\ref{sec:comparison_Xray_catalogues} shows a comparison of the eRASS:3 catalogue with the eRASS1 catalogue \citep{Merloni2024}, as well as with that of the all-sky ROSAT survey \citep[2RXS,][]{Boller2016}, the CSC\,2.1 from {\it Chandra} \citep{Evans2024}, and the 5XMM-DR15 from XMM-{\it Newton} (Webb et al., in prep.) serendipitous catalogues. Section~\ref{sec:CTP_class} presents the multi-wavelength counterparts of the eRASS:3 X-ray sources. Section~\ref{sec:dr2_description} describes the content of eROSITA-DE DR2. We draw our conclusions in Sect.~\ref{sec:summary_outlook}.

%
%

\begin{figure}
    \centering
    \includegraphics[width=\columnwidth]{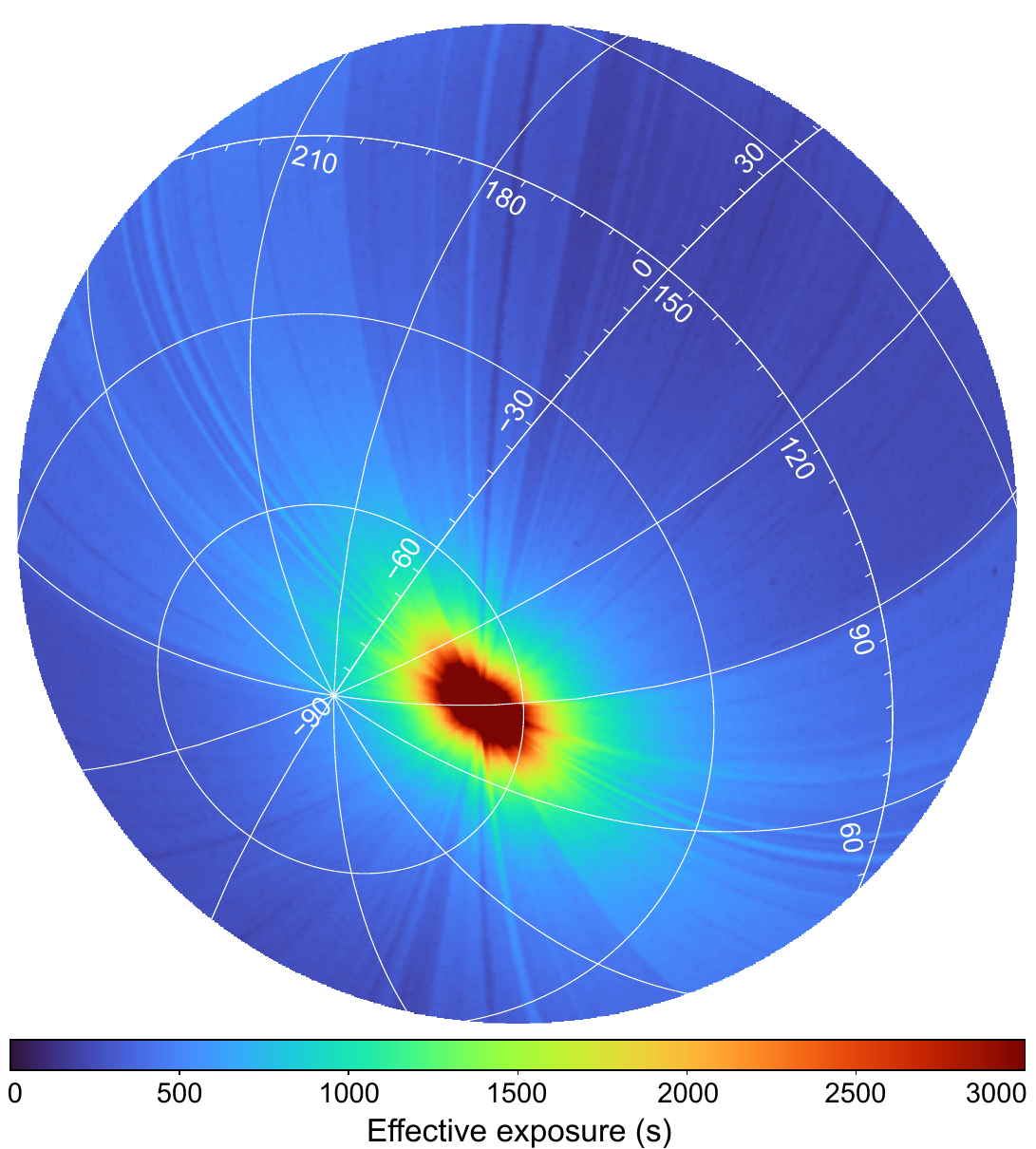}
    \caption{Effective (vignetted) eRASS:3 exposure map (J2000, zenithal equal area projection). The values in the map show the exposure time multiplied by the average of the ratio of the (vignetted) effective area to the on-axis effective area in the energy band $0.6-2.3$~keV. Only the eROSITA-DE (western) Galactic hemisphere is displayed. The highest exposure is reached toward the South Ecliptic Pole, where the map is saturated (see Sect.~\ref{sec:operations} for details).}
    \label{fig:erass3_exposuremap}
\end{figure}

\section{All-sky survey observations}
\label{sec:operations}
The SRG observatory performs an all-sky survey by constantly rotating around an axis pointed near the Sun. Each SRG revolution lasts four hours, thus scanning the sky at a rate of 0.025~deg~s$^{-1}$, while its rotation axis slowly shifts by $\sim1^\circ$ per day to follow Earth's orbit. This pattern allows eROSITA to cover the entire sky every six months, with each sky point typically observed six times, each for up to 40 seconds \citep{Predehl2021,Merloni2024}.

In brief, eRASS1 was conducted from December 12, 2019, to June 11, 2020, spanning 184 days; eRASS2 took place immediately afterwards, until December 15, 2020 (188 days); and eRASS3 lasted until June 16, 2021 (184 days). During this time, daily ground contacts were conducted uninterrupted, and a total telemetry volume of $\sim242$~GB ($\sim435$~MB/day) was downlinked. A timeline of the most significant operations milestones during eRASS2 and eRASS3 is presented in Table~\ref{tab:erass23_operations}. Figure~\ref{fig:erass3_exposuremap} shows the combined half-sky exposure map of the eROSITA-DE hemisphere during eRASS:3 in the $0.6-2.3$~keV energy band. Slight inhomogeneities in sky coverage arise from the spacecraft’s halo orbit around L2 and from the adjustments needed to maintain Earth within the downlink antenna's cone. Regions near the ecliptic poles receive longer exposure due to the scanning strategy (i.e. all scan circles intersect there). Effective exposure values range from ${\sim}140$\,s at the ecliptic equator to more than \num{70000}\,s close to the ecliptic poles (not visible in the colour scale of Fig.~\ref{fig:erass3_exposuremap}).

\begin{figure}
    \centering
    \includegraphics[scale=0.365]{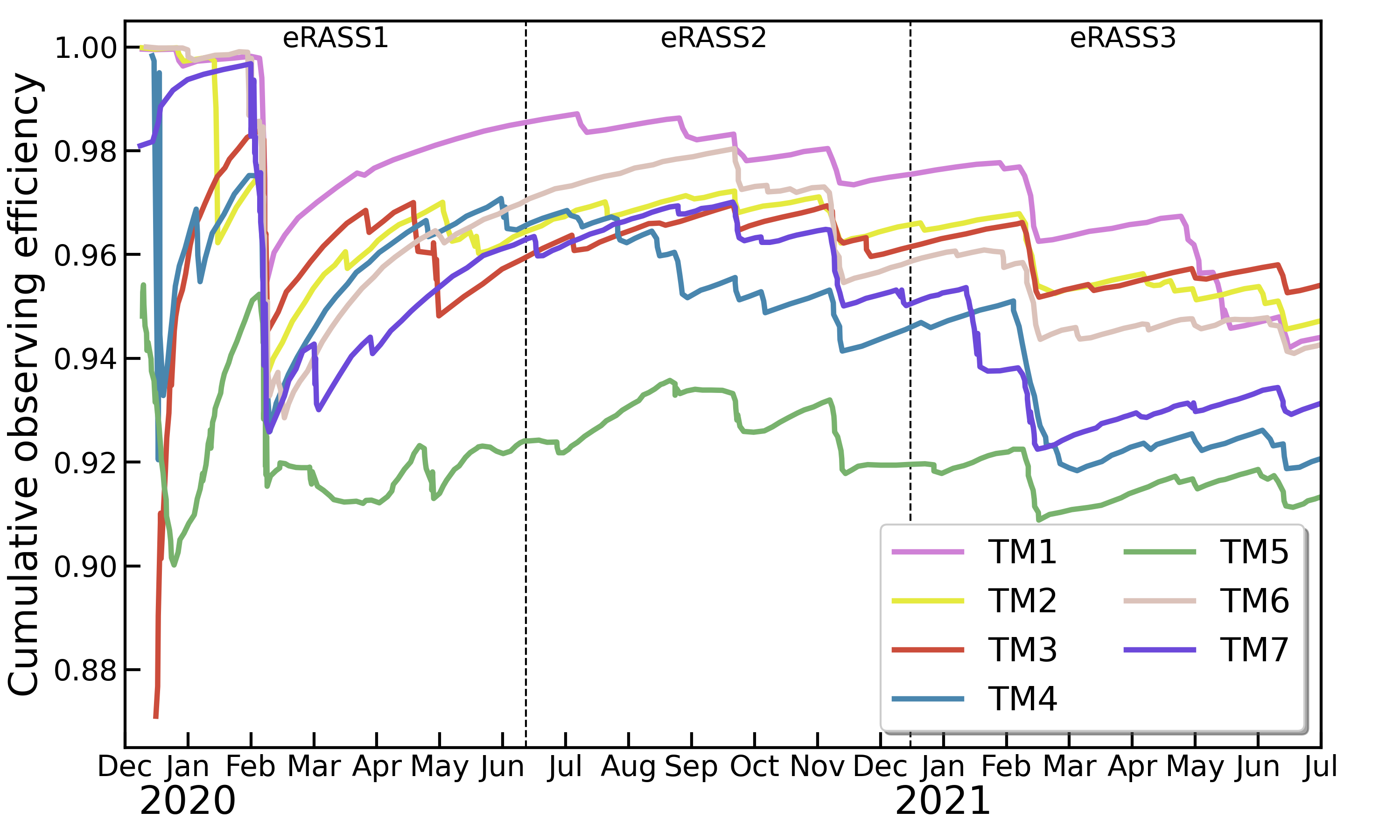}
    \caption{Cumulative observing efficiency (as a percentage of the elapsed survey time) as a function of time for each eROSITA camera (labelled according to the associated Telescope Module, TM) during eRASS1 to eRASS3.}
    \label{fig:erass1to3_obseff}
\end{figure}

The Mission Control Centre in Moscow used deep-space antennas at Ussuriysk and Bear Lakes for data downlink. eROSITA achieved an average observing efficiency of $\sim94\%$ (see Fig.~\ref{fig:erass1to3_obseff}), with brief losses due to single event upsets (SEUs) affecting the camera electronics (CE) and the Interface and thermal controller (ITC), as well as light leaks in telescope modules 5 and 7 \citep[for details, see][]{Coutinho2022,Predehl2021}. During eRASS3, eROSITA experienced at least four micrometeoroid impacts, most notably a major hit on 23 February 2021 that damaged over 6\,000 pixels on telescope module 4 (TM4) \citep[see][]{Freyberg2022,Coutinho2022}. These pixels produced spurious events, appearing bright in images and sometimes saturating telemetry. Affected pixels were flagged as bad on board, while orthogonally adjacent pixels were also disabled for the spectroscopy. To mitigate the telemetry impact, individual energy thresholds were raised for moderately affected pixels, thereby reducing false-event rates to normal levels. This treatment reduced the number of bad pixels on TM4 from over 6\,000 to fewer than 300. The CE onboard processor enabled this pixel-level adjustment, significantly improving data quality within telemetry limits. After applying the mitigation procedures, no residual signatures of micrometeoroid impacts remain in the data and, consequently, in the X-ray source catalogues presented in this work. Aside from these impacts and the short-lived SEU events, all eROSITA subsystems remained fully operational during eRASS1-3, with no lasting damage.

Routine filter wheel closed (FWC) observations, conducted every seven days per camera \citep[see][]{Merloni2024} to monitor particle background levels, were halted for most cameras in July 2020. This change was prompted by an observed increase in background noise whenever the filter wheel was in motion. It typically took about 10 minutes after the wheel stopped for the background to return to nominal levels. In addition to the $18$ short FWC observations, around $8$ were performed for each camera in eRASS2. An exception was TM4, which did not exhibit a significant background increase when its filter wheel moved; thus, FWC observations for TM4 continued through eRASS2 and 3, contributing an additional 40 FWC observations.

%
%

%
\section{eROSITA calibration and data processing}
\label{sec:cal_processing}
This section provides a concise summary of the eROSITA calibration and data processing procedures. The eRASS:3 data were processed using version 030\footnote{This software version is not made public, as the present release includes only the higher-level products, not the raw data; see Sect.~\ref{sec:dr2_description}. The software used for DR2 is not validated to run on the DR1 or EDR data.} of the eROSITA Science Analysis Software System (eSASS) pipeline, which includes several calibration and processing improvements over version 010 (used in DR1) and version 020, which was used for later work based on proprietary eROSITA data \citep[see][for further details]{Merloni2024}. The standard pipeline structure and its modules remain unchanged and are described in detail in \cite{Brunner2022} and \cite{Merloni2024}. 

Compared to the processing used for DR1, version 030 introduces refinements to the energy calibration, astrometric calibration, event reconstruction, flare filtering, and source detection procedures. For catalogue users, the main practical consequences are a recovery of the small sensitivity loss present in earlier processing versions, improved astrometric accuracy through the application of time-dependent boresight corrections, and more robust source detection through the exclusion of flare-contaminated data. Overall, the changes relative to DR1 are generally small, affecting source parameters and catalogue content at a low level. The main changes are described in the following subsections. The versions of the eSASS tasks and calibration database used in this work are listed in Appendix~\ref{appendix:esass}. Table~\ref{tab:esass_versions} summarises the updates introduced in eSASS version 030, while Table~\ref{table:caldb} provides the corresponding calibration file versions.

\subsection{Calibration}
\label{sec:calibration}
As in processing versions 010 and 020, version 030 continues to use the ground-based point spread function (PSF) calibration, whose reliability has been demonstrated by \cite{Merloni2024}. Further refinements of the in-flight PSF and vignetting calibrations are ongoing, but are not expected to produce substantial changes in source positions or fluxes. In particular, the unreleased vignetting model reduces the survey-averaged effective area above $4$~keV by approximately $15-20\%$, while changes in the primary $0.2-2.3$~keV energy band are typically only a few percent. Consequently, the impact on the soft-band and broadband source fluxes is expected to be small for most sources. The following subsections summarise the main calibration and processing updates introduced in version 030.

\subsubsection{Energy calibration}
\label{sec:energy_calibration}
The energy calibration in version 020 was based on a combination of ground-based measurements and in-flight $^{55}$Fe exposures obtained in two calibration runs in November-December 2019 and November 2020, from which the temporal charge transfer inefficiency (CTI) increase was derived and taken into account as a constant linear trend. This approach was supported by a third calibration run in February 2021, which showed that the temporal increase in CTI continued at the same rate. However, two subsequent calibration runs in August 2021 and February 2022 exhibited deviations from extrapolating the previously observed linear trend, motivating a refinement of the energy calibration by utilising all available calibration data. In addition, the now-available time span of more than 2 years of in-flight calibration data enabled the determination of temporal trends in the effective gain (caused by CTI changes in the framestore area). All these changes were considered individually for each CCD column for the reconstruction of the photon energies in the 030 data processing. Several tests were performed to demonstrate the accuracy of the updated energy calibration.

In the first test, the new calibration was applied to the $^{55}$Fe exposures obtained in all the five calibration runs between Nov 2020 and Feb 2022, showing that the absolute energy scale at Al--K ($1.486$~keV) could be reconstructed to an accuracy of $\pm1\mbox{eV}$, for each of the five TMs which were not affected by the light leak, despite CCD temperature changes and the CTI increase due to radiation damage. The improved correction of CCD-column-specific temporal CTI changes minimised the variations in the absolute energy scale between individual CCD columns and reduced the temporal full width at half maximum (FWHM) increase over the full CCDs by several eV (e.g. by $\sim8\mbox{ eV}$ at 5.9~keV in the Feb 2022 exposure) compared to the 020 processing.

\begin{figure}
    \centering
    \includegraphics[width=0.495\textwidth]{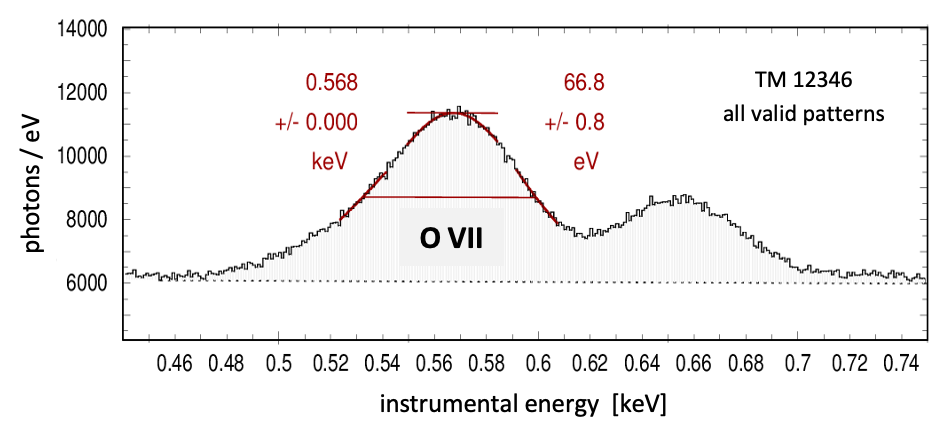}
    \caption{Spectrum of the O~VII/O~VIII region, obtained by stacking the energies of all valid pixel patterns in TM1-4 and TM6 and the full FoV. The dashed line indicates the empirically determined background level. After subtracting the background, three Gaussian functions were fitted to the peak and the left and right wing regions (red curves). Two straight red lines mark the peak flux and half its value. The intersection of the lower line with the red curves yields the FWHM, which is 2.355 times the $\sigma$ of the line width. The statistical $1\sigma$ error of the peak position is less than 1~eV.}
    \label{fig:stacked_spectrum}
\end{figure}

A second test took advantage of the fact that the diffuse X-ray sky exhibits pronounced emission from O~VII ions, consisting of three main components at $0.561$, $0.569$, and $0.574$~keV ~\citep{Ponti2023}. Although the relative flux between these components shows some spatial and temporal variations, the diffuse O~VII emission can serve as a proxy to assess the accuracy of the energy calibration over more than two years. Such long-term monitoring is particularly important because the CCD temperatures varied between $-86$~C and $-79$~C during this period (with similar ranges for all TMs). For this test, the TM1-4 and TM6 data, taken between Dec 2019 and Feb 2022 (eROdays\footnote{eROday = time interval of 4-hours, see Sect.~\ref{sec:dataprocessing}.} $43850-48550$), were sampled in steps of $100$~eROdays and processed with the updated energy calibration. The stacked spectrum (TM1-4 and TM6, all valid patterns, full field of view -- FoV) showed an emission line at $0.568$~keV with a FWHM of $66.8\pm0.8\mbox{ keV}$ (see Fig.~\ref{fig:stacked_spectrum}). For comparison, the ground-based calibration measurements of the O-K emission line at $0.527$~keV, obtained in 2016 at CCD temperatures between $-95.8$ and $-94.2$~C, yielded a FWHM of $64\pm0.1\mbox{ eV}$. If one considers that, compared to the ground-based measurements, the in-flight measurements $i)$ were performed over a time span of more than two years, $ii)$ over a wider range of CCD temperatures, $iii)$ at $10-15$~C higher CCD temperatures, $iv)$ with CCDs subject to increasing CTI due to radiation damage, $v)$ at a somewhat higher photon energy, $vi)$ did not measure a sharp emission line, but an emission complex consisting of three major components, and $vii)$ that the diffuse emission regions are not exactly at rest relative to SRG/eROSITA, adding some further broadening, then the in-flight FWHM of $66.8$~eV offers an extremely close comparison with the ground-based FWHM of $64.0$~eV.

A third test investigated the pointed observations of the supernova remnant 1E\,$0102.2-7219$, which exhibits a rich spectrum of narrow emission lines and is regarded as a standard calibration source \citep{Plucinsky2017}. This source was observed three times, in Nov 2019, Jun 2020, and Nov 2021. It turned out that, in all cases, the spectra can be well reproduced by the model spectrum recommended by IACHEC\footnote{\url{https://iachec.org/}}, with only two free parameters: the global normalisation and an energy offset. The derived offsets indicate that the absolute energy scale can be determined to an accuracy of approximately $\pm3\mbox{ eV}$. In this context, the Nov 2021 spectrum is particularly remarkable, because the CCD temperatures increased by $+8~^\circ\mbox{C}$ during that observation. These tests clearly demonstrate the high accuracy of the time-dependent energy scale calibration.

\subsubsection{Recovery of lost events}
In processing version 010, an attempt was made to improve the spectral resolution homogeneity of the CCDs. The spectral resolution depends on the detectors' low-energy threshold, which is required to account for detector noise. Prior to launch, the noise properties of each pixel were determined and used to derive pixel-specific low-energy thresholds, thereby evenly distributing the available telemetry across all pixels. As a first step towards establishing a homogeneous spectral resolution, a somewhat higher, but constant threshold ($46$ in analogue-to-digital units, adu) was applied in the very first (001) processing of the TM1-4 and TM6 data for eROSITA-DE Early Data Release (EDR\footnote{\url{https://erosita.mpe.mpg.de/edr/}}). This, however, was only an approximation because $i)$ the conversion from adu to eV varies across CCD columns, and $ii)$ due to CTI losses, the adu values decrease with increasing distance from the readout node for a given incident photon energy. These effects have now been taken into account since version 010, where constant thresholds in eV are applied. To maximise spectral resolution, they were chosen as low as the original thresholds applied on board allowed. The resulting values are 50 eV (TM1), 60 eV (TM2), 60 eV (TM3), 55 eV (TM4), 105 eV (TM5), 50 eV (TM6), and 125 eV (TM7). The same values apply to versions 010 and 020. In version 030, they were increased to $60$, $70$, $70$, $65$, $105$, $60$, $125$~eV, respectively, as it turned out that, in the 010 and 020 versions, noise events had made some valid pixel patterns invalid, causing a $\sim3.8\%$ reduction in sensitivity in TM1-4 and TM6. A detailed analysis\footnote{See \url{https://erosita.mpe.mpg.de/dr1/eROSITA\_issues\_dr1/threshold\_issue\_dr1.html}} showed that raising thresholds by just $10$~eV is sufficient to mitigate these problems, recovering the loss in sensitivity.

\subsubsection{Time corrections}
The data in eROSITA CCDs are read periodically and packaged into so-called science frames. Although the time between CCD readouts (approximately $50$~ms) is governed by a very regular clock in the camera electronics, the precision of this interval (which is of the order of $10^{-6}$) is not preserved in the telemetry. Frame timestamps are instead generated from a counter with a precision of $\sim10$~ms. eSASS task {\tt frameprep} (see Sect.~\ref{sec:dataprocessing}) reconstructs the original values by a two-step process of first establishing the sequence of frames (which involves the determination of the size of any gaps in the sequence), then fitting a model of the true CCD readout times to the coarse timestamp data. The result is stored in the TIME column of the event list. Average uncertainties in the science-frame timestamps are thereby reduced from $\sim5$~ms to $\sim30~\mu$s.

We note that corrections have not yet been performed for {\it i}) the constant offset between the frame integration centre time and the time at which the timestamp is applied; {\it ii}) time-model discontinuities between adjacent eROdays; and {\it iii}) drift (about $\sim13$~ms per day, with resets before the total exceeds $1$~s) in the main spacecraft clock. Therefore, timing uncertainties in photon arrival measurements can occasionally reach $1$~s. This $1$~s uncertainty only affects event-based analysis; it does not affect the eRASS:3 source catalogue presented in this work.

\subsubsection{Astrometric corrections: Boresight correction}
\label{sec:tel_astrometry}
Astrometric corrections were applied to the eRASS:3 catalogues after merging the detections from the individual sky tile catalogues. The corrections were determined by matching point-source detections with mid-infrared counterparts from the AllWISE catalogue \citep{Cutri2021}. For each of the six monthly all-sky surveys,$~1.4\times 10^{6}$ AllWISE matches were found. Correction values were determined by computing the median offsets from the counterpart positions in ecliptical coordinates $\lambda$,~$\beta$ (ecliptic longitude and latitude), within stripes of 1 degree width in ecliptic longitude $\lambda$. The astrometric correction values determined for the eRASS1-eRASS5 catalogues created with the previous pipeline version 020 were mapped to a time series using the SRG attitude data. This time series covers eRASS1-5 and is stored in the calibration file {\tt srg\_tposcorr\_190611v01.fits}. In the current pipeline version 030, this calibration file was used to apply astrometric corrections to the event positions.

The implementation of the time-dependent boresight calibration in version 030, together with improved source-detection convergence, has significantly reduced the systematic positional uncertainty. Averaging source positions across three surveys, compared with a single survey in DR1, likely provides an additional contribution. Overall, the systematic positional error was reduced from approximately $0.9\arcsec$ in DR1/eRASS1 to $0.4\arcsec$ in DR2/eRASS:3.

\subsection{Data processing}
\label{sec:dataprocessing}
The eROSITA telemetry data were transmitted in real time to the SRG operations centre and the Space Research Institute (IKI), where they were stored and forwarded to the Max-Planck Institute for Extraterrestrial Physics (MPE) via a data exchange server. These files were processed into FITS format and passed to two pipelines: the pre-processor archiver and the Near Real Time Analysis (NRTA). Data are organised into 4-hour intervals called ‘eROdays’ that match SRG’s survey rotation, which are archived upon completion, thereby triggering further processing. The NRTA pipeline enables rapid monitoring of instrument health and alerts for transient events, while archived data proceeds to standard analysis \citep[see Appendix C of][]{Merloni2024}. As in previous all-sky survey processings, the sky is divided into $4\,700$ unique, overlapping sky tiles of $\sim8.78$~deg$^{2}$, arranged into $61$~declination zones. Each sky tile is centred within a larger $3.61^\circ\times3.61^\circ$ field that overlaps with neighbours by $15\arcmin - 18\arcmin$ to minimise edge effects during source detection.

The sky tiles of eRASS:3 data were processed at MPE using the eROSITA standard pipeline version 030, which includes modules for event handling (TEL chain), event file and image creation (EXP chain), exposure and background map generation and source detection (DET chain), and the production of source-specific products (SOU chain). These processing chains utilise tasks from the eSASS. For a detailed description of the eSASS tasks, the associated eROSITA calibration database, and the standard processing pipeline, see \cite{Brunner2022} and \cite{Merloni2024}. In the following, we highlight improvements in data processing and changes to eSASS tasks relative to versions 010, used in DR1, and 020, used for proprietary data analysis. Appendix~\ref{appendix:esass} presents a list of the tasks for which a functional change has occurred in 030 processing, as well as the calibration files that have been updated with respect to previous releases.

\subsubsection{TEL chain}
The TEL chain handles event file preparation, pattern reconstruction, energy calibration, and attitude calculation. The data processing workflow in 030 has been slightly updated. Functionality previously handled entirely by the task {\tt evprep} is now split between a new task, {\tt frameprep}, and a simplified {\tt evprep}. {\tt frameprep} performs a more thorough evaluation of science frame quality, and fits a time model to correct for truncation jitter in science-frame timestamps (see Sect.~\ref{sec:calibration}). Introduced in data processing version 030, the task {\tt hotpixfind} replaces {\tt ftfindhotpix}\footnote{{\tt ftfindhotpix} was not used in previous pipeline configurations.}. Although {\tt hotpixfind} can detect overactive pixels and flag them in the event list's {\tt BADPIX} table, this functionality is under development and was not enabled in the 030 pipeline. In 030, the task was limited to copying relevant entries from the calibration database's bad pixel component. As mentioned in Sect.~\ref{sec:calibration}, the energy and flux reconstructions have been improved, and time-dependent boresight correction has been applied in tasks {\tt pattern}, {\tt energy}, and {\tt evatt}, respectively.

\subsubsection{EXP chain}
In 030 processing, the event lists generated for each sky tile exclude events with invalid patterns, those outside the FoV, those outside good time intervals (GTIs) due to defective frames, and data artefacts and corrupted frames. The task {\tt flaregti} no longer writes GTI extensions directly to the event files; instead, it produces a {\tt FlareGTI} file containing a single {\tt STDGTI} extension that encodes the GTI common to all TMs. The pipeline-generated images are created with flare-cleaned GTIs, ensuring that, unlike in the 010 and 020 processings, source detection is performed on data free of background flares. The fraction of observing time affected by soft-proton flares amounts to $1.35\%$, $2.33\%$, $1.49\%$, $1.73\%$ for eRASS1, eRASS2, eRASS3, and eRASS:3, respectively. This demonstrates that flare-induced data losses are minimal and do not significantly compromise the quality of the eROSITA observations.

\subsubsection{DET chain}
\label{subsec:det_chain}
This module creates exposure and sensitivity maps, performs source detection, and extracts aperture photometry on the detected sources. The source detection tasks have remained stable across processing versions; a full description is provided in \cite{Brunner2022} and \cite{Merloni2024}. We just emphasise that, for both point and extended sources, the rate, count, and flux values reported in the eRASS catalogues are derived from the scaling of the best-fit PSF or PSF-folded extent model. These quantities, by definition, represent integrals to infinite radius, although the model fits are performed within circular sub-images of $1\arcmin$ radius. Finally, for each sky tile, two source lists are produced: a single-band catalogue (‘1B’) based on detections in the broad $0.2-2.3$~keV energy band for maximum sensitivity, and a three-band catalogue (‘3B’; $0.2-0.6$, $0.6-2.3$, and $2.3-5.0$~keV), where detection is performed simultaneously across bands and both single-band and combined likelihoods are computed for each source.

\subsubsection{SOU chain}
As in the 010 and 020 processes, in the 030 process, source products (spectra, background spectra, response matrices, ancillary response files, and light curves) were created using the {\tt srctool} task for the subset of bright sources with a detection likelihood greater than 20 in the single-band 1B catalogue.

\begin{table}
\caption{Number of sources flagged as potentially spurious.}
\label{tab:source_flag_summary}
\renewcommand{\arraystretch}{1.2}
\begin{tabular}{lrrrr}
\hline
\hline
Flag & {\it M} PS & {\it M} EXT & {\it H} PS & {\it H} EXT \\
\hline
\texttt{FLAG\_SP\_SNR} & 6\,569     &  4\,542  &  63  & 51 \\
\texttt{FLAG\_SP\_BPS} & 3\,069     &  1\,443   & 69  & 99 \\
\texttt{FLAG\_SP\_SCL} & 4\,060     &   415   & 57  & 8 \\
\texttt{FLAG\_SP\_LGA} & 569      &   452   & 21  &  7\\
\texttt{FLAG\_SP\_GC\_CONS} & 6\,921 &  3\,641  & 98  & 298 \\
\hline
Any SP Flag & \num{21165} & \num{10480} &  305 & 461\\
\hline
\texttt{FLAG\_NO\_RADEC\_ERR} & 5\,261 & 1\,566  & 20 &  34 \\
\texttt{FLAG\_NO\_EXT\_ERR}   &    0 & 1\,290 &  0 &  33 \\
\texttt{FLAG\_NO\_CTS\_ERR}   &  5\,185 & 1\,312 &  7  &  1\\
\hline
\end{tabular}
\tablefoot{`PS' stands for point sources, `EXT' for extent-selected sources, and we separate the main (`{\it M}') and the hard (`{\it H}') catalogues. The `Any SP Flag' row indicates the number of sources that are flagged by any of the five identifiers of potential spurious sources in overdense regions (see Sect.~\ref{sec:spurious_flagging}). The last three lines give the number of sources for which the statistical error estimate from the pipeline failed (see Sect.~\ref{sec:known_issues} for more details).}
\end{table}

\subsubsection{Known data processing issues}
\label{sec:known_issues}
The problems of poor convergence of parameters of bright sources, uncertainty estimations on the position, counts, and extent of faint sources, as well as the inclusion of the leap seconds discussed in \cite{Merloni2024}, have been addressed in the 030 processing data (see Appendix~\ref{appendix:esass}). We refer to Sect.~$4.5$ of \cite{Merloni2024} for details on the still-known issues of eROSITA data processing: pile-up in bright sources, `event quota' mechanism due to telemetry constraints, optical light leak contamination in TM5 and TM7. Due to remaining issues with PSF fitting convergence and possible event pile-up, it is advisable to allow an additional systematic error of the order of 1 arcsec for very bright sources ($> 1000$ source counts).

In some cases, the source-detection algorithm failed to estimate errors, leaving certain sources without reliable uncertainties in position, counts, or extent. This primarily affects low-probability or spurious detections in regions of extended emission, although some significant detections are also impacted. These sources are flagged in the catalogue as {\tt FLAG\_NO\_RADEC\_ERR}, {\tt FLAG\_NO\_CTS\_ERR}, and {\tt FLAG\_NO\_EXT\_ERR} (Sect.~\ref{sec:spurious_flagging} describes in detail these spurious source flags). Table~\ref{tab:source_flag_summary} lists the number of sources flagged in each category in the eRASS:3 catalogues.

\begin{figure*}
    \centering
    \includegraphics[width=0.495\textwidth]{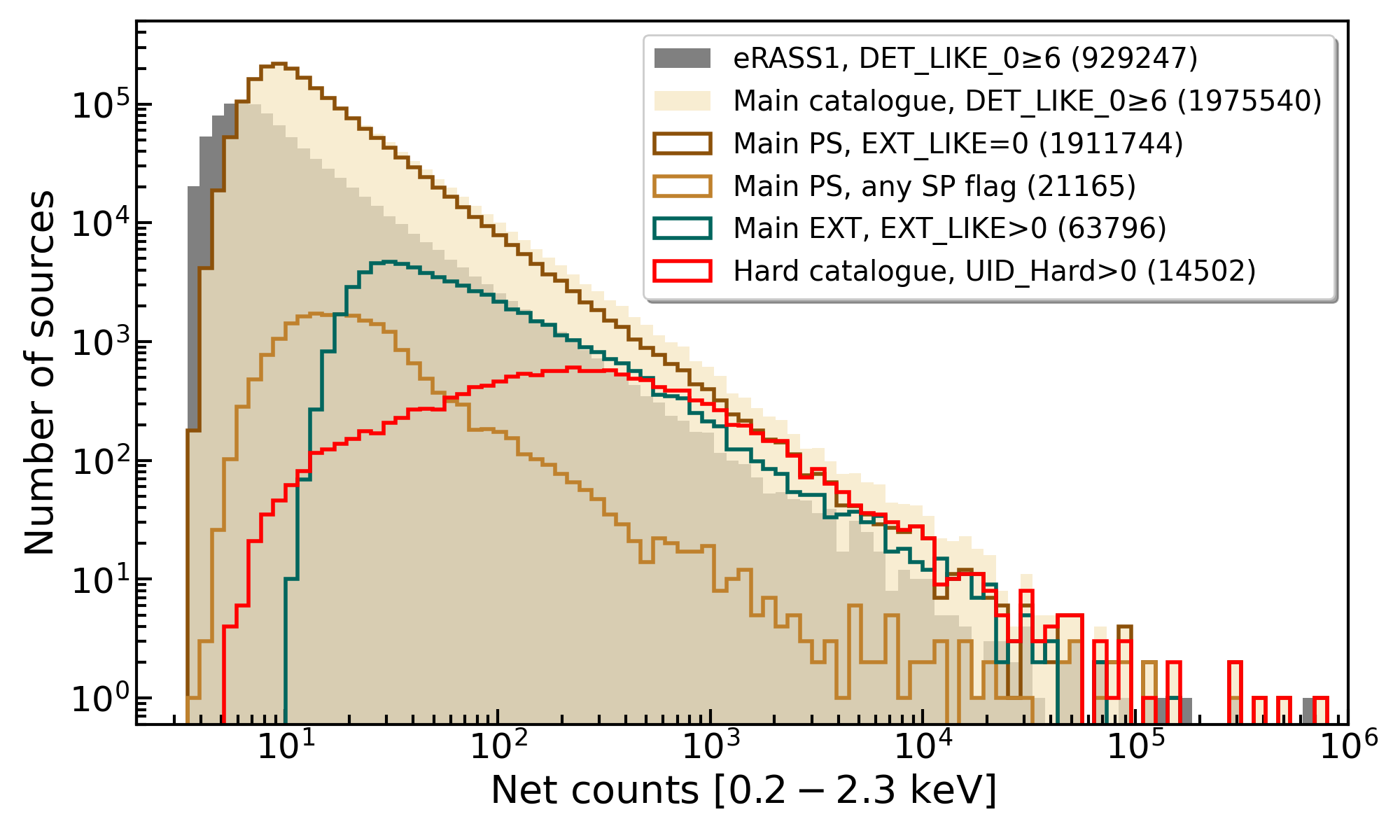}
    \includegraphics[width=0.495\textwidth]{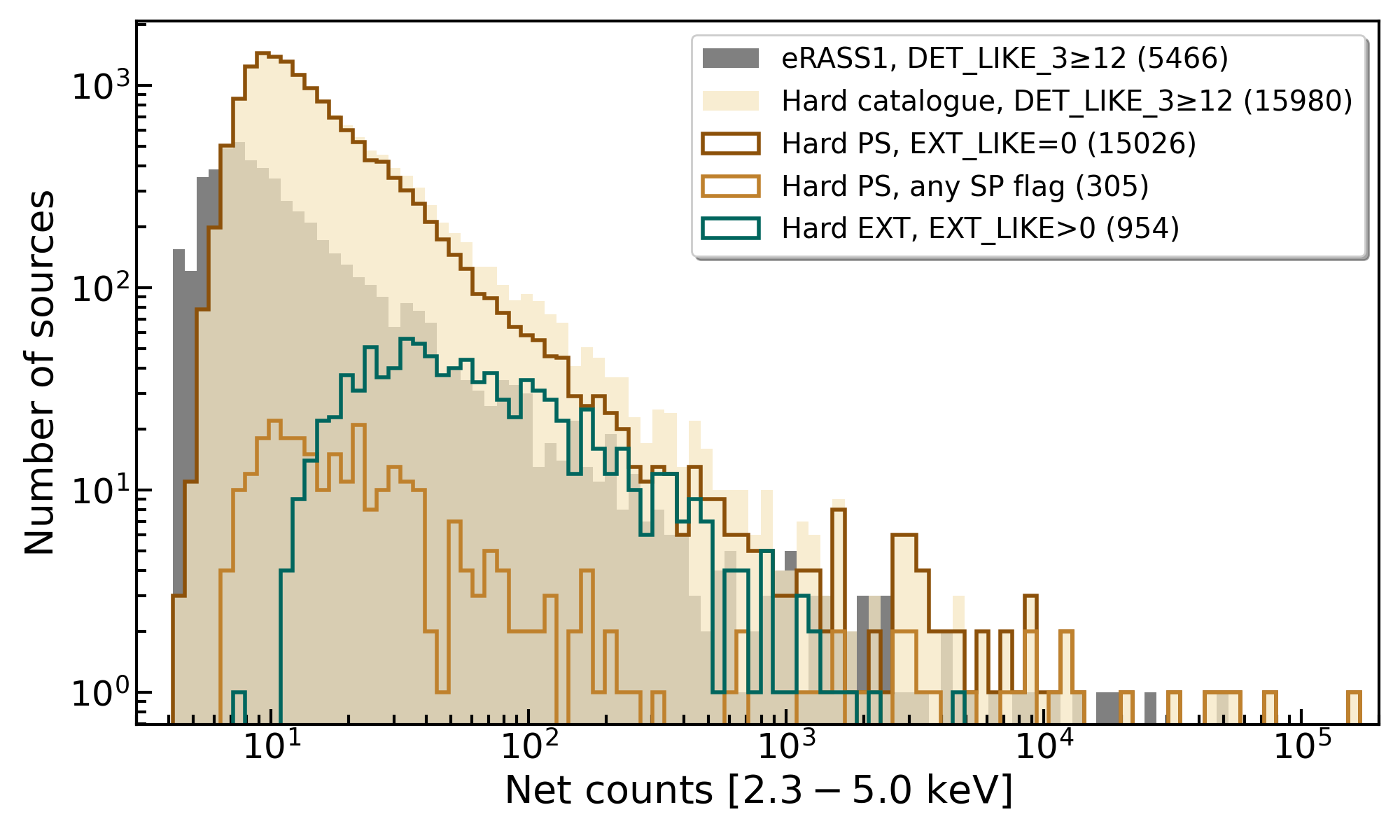}
    \includegraphics[width=0.495\textwidth]{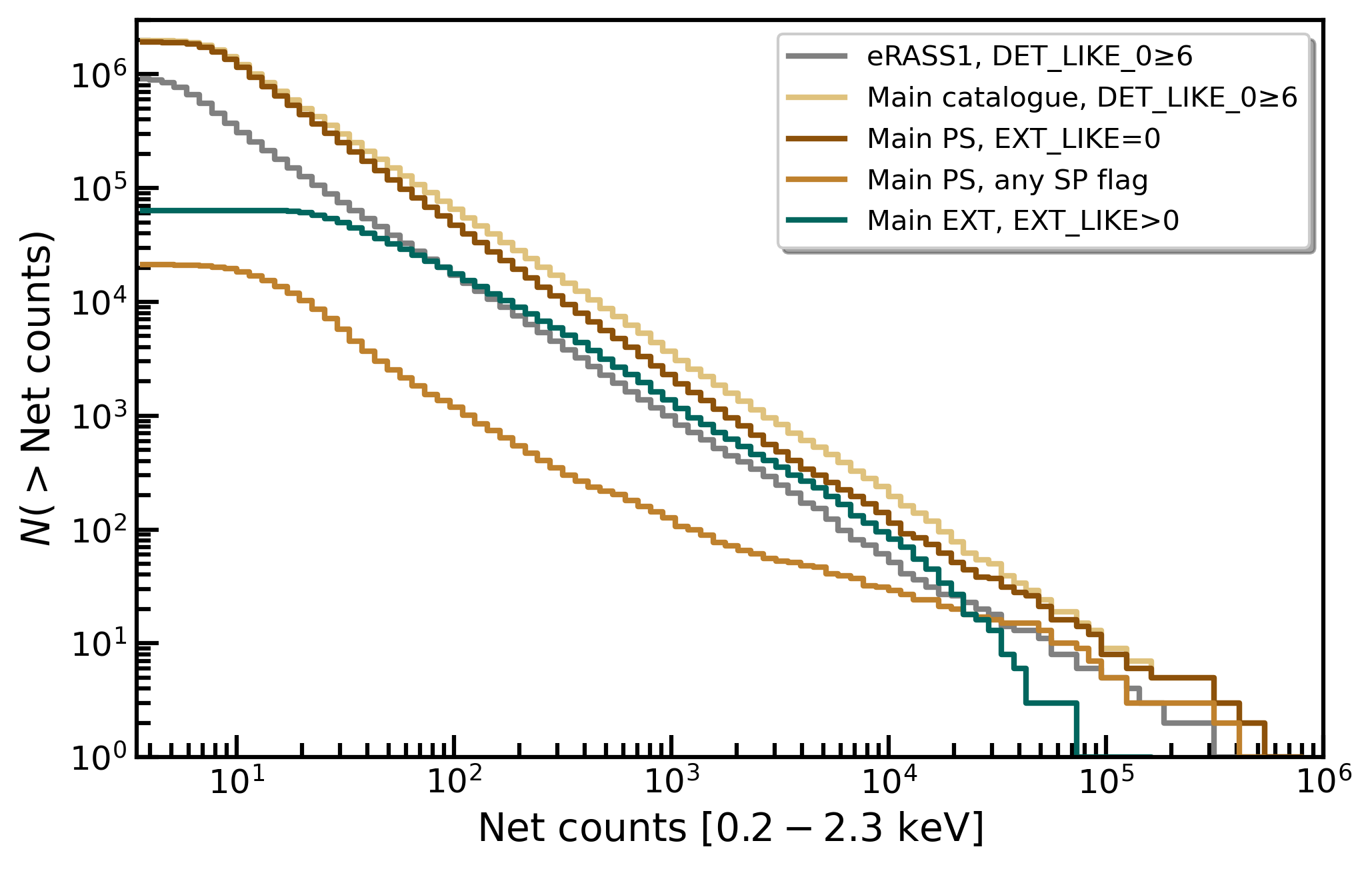}
    \includegraphics[width=0.495\textwidth]{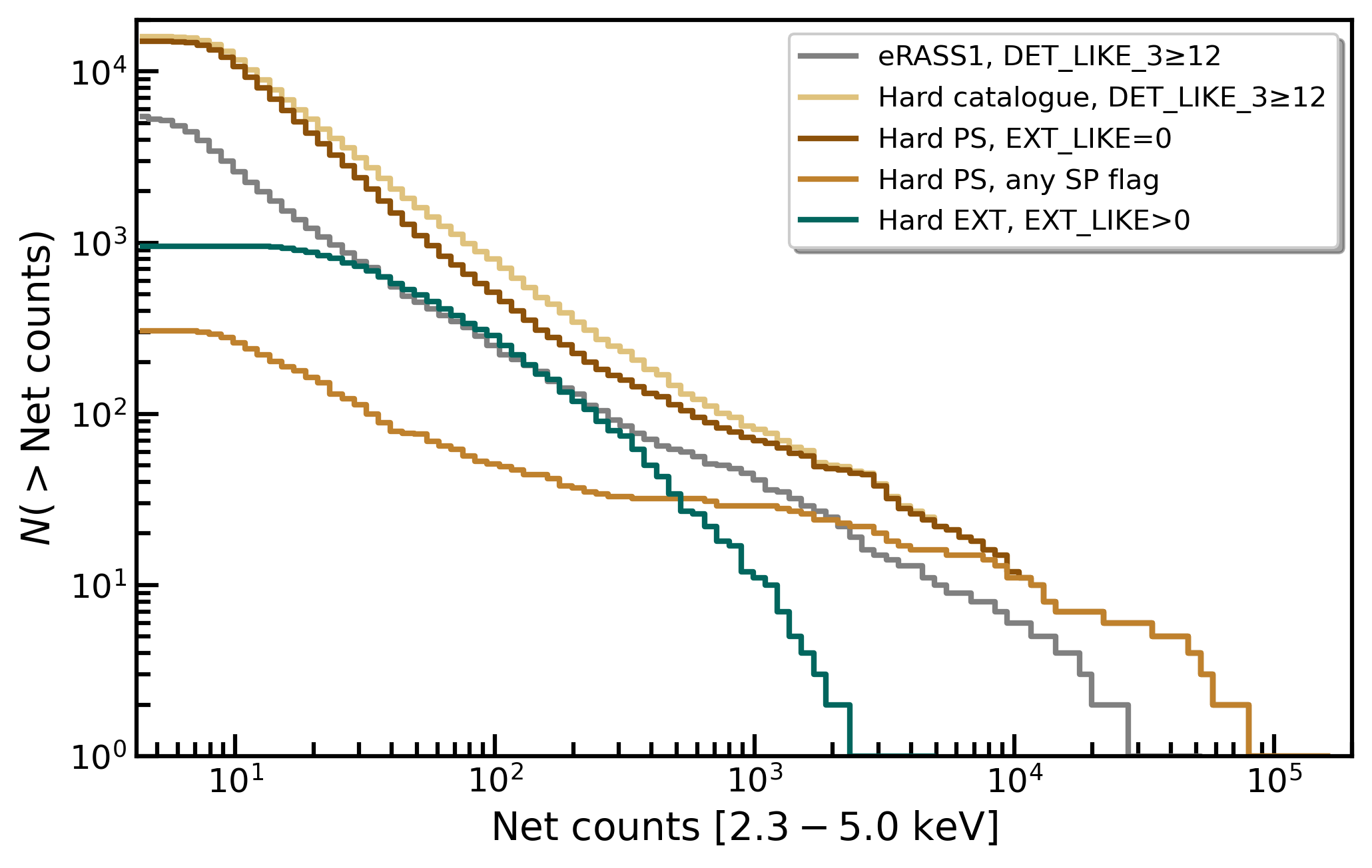}
    \caption{{\it Top panels:} Distribution of net counts for the eRASS:3 main ({\it left}) and hard ({\it right}) catalogues, shown in yellow. {\it Bottom panels:} Reverse cumulative distributions, i.e. the number of sources with net counts above a given value, for the same samples. In all panels, point sources (PS; {\tt EXT\_LIKE = 0}) and extended sources (EXT; {\tt EXT\_LIKE > 0}) are shown in brown and green, respectively. Point sources flagged with any SP flag (see Table~\ref{tab:source_flag_summary}) are displayed in gold. The grey histograms indicate the corresponding distributions from the eRASS1 catalogues for comparison. Sources detected in both the eRASS:3 main and hard catalogues are shown in red in the top-left panel.}
    \label{fig:net_counts_distribution}
\end{figure*}

After the 030 processing, an issue was identified affecting the bad-pixel calibration files for TM4, TM5, and TM7, starting from specific times unique to each telescope module. The structure of these calibration files was modified at that point; however, the new format was incompatible with the processing task's requirement to apply bad-pixel filtering to the event lists. As a result, bad pixels were not fully excluded from the data thereafter. Additionally, the onset time of a known camera malfunction was incorrectly reported. This led to the persistence of bad pixels in the eRASS2 dataset, which subsequently propagated into eRASS:3. The presence of these unfiltered bad pixels resulted in spurious sources in the sky images. These spurious sources were later identified and removed during the construction of the final source catalogues.

Finally, in several cases, significant sources were found to be missing from the catalogues. These sources are typically located in the vicinity ($\sim 1\arcmin$) of other similarly bright sources and were filtered out during the {\tt erbox} peak finding stage. Based on the comparison of the eRASS:3 main catalogue with the XMM-{\it Newton} 5XMM-DR15\footnote{http://xmmssc.irap.omp.eu/Catalogue/5XMM-DR15/5XMM\_DR15.html} (Webb et al., in prep.) catalogue (see Sect.~\ref{sec:comparison_Xray_catalogues}) we find $8$ sources, brighter than $10^{-14}$~erg~cm$^{-2}$~s$^{-1}$ (in the $0.2-2.0$~keV energy band) in 5XMM-DR15, which could be visually confirmed in eRASS:3 images, but were not detected or had large positional offsets in eRASS:3. Given the $ 4\% $ coverage of 5XMM-DR15 in the eRASS:3 footprint, we estimate that significant sources of the order of $200$ might go undetected due to this issue in the eRASS:3 main catalogue.

Since the discovery of X-rays from comets \citep{Lisse1996,Dennerl1997}, it has been recognised that observations of the X-ray sky are affected by diffuse foreground emission arising from solar-wind charge exchange (SWCX) between highly ionised heavy ions in the solar wind and tenuous neutral gas in the solar system \citep{Cravens1997}. Such gas is present both around the Earth (the geocorona) and around the Sun (the heliosphere). While eROSITA is not affected by geocoronal SWCX, heliospheric SWCX contributes with a detectable, time-variable foreground component in eRASS data, particularly from eRASS1 onward during the rise phase of the solar cycle. This emission predominantly affects energies below 1~keV and manifests as a spatially smooth but temporally variable background component. Its presence must be accounted for in analyses of extended sources and diffuse X-ray emission. For unresolved sources, heliospheric SWCX results in reduced effective sensitivity below 1~keV due to the elevated background. Systematic effects can be mitigated by subtracting a locally determined background. In general, emission at energies above 1~keV is not affected by SWCX. A detailed analysis of SWCX in eRASS:5 is presented by \cite{Dennerl2026}.

%
%

%
\section{The eRASS:3 X-ray catalogues}
\label{sec:erass3_catalogs}
We provide two distinct X-ray source catalogues of the western Galactic hemisphere for eRASS:3: {\it i}) the main catalogue, containing sources detected in the $0.2-2.3$~keV energy band via the 1B detection process, and {\it ii}) the hard catalogue, containing sources detected in the $2.3-5.0$ keV band via the 3B detection process (see Sect.~\ref{sec:dataprocessing}). This follows the same methodology established for the performance verification eFEDS survey \citep{Brunner2022} and later applied to the eRASS1 catalogues \citep{Merloni2024}. In the following sections, we present the construction procedure and overall properties of the catalogues.

\subsection{Catalogue creation and astrometric corrections}
The source lists produced for each sky tile by the DET chain (see Sect.~\ref{subsec:det_chain}) are merged into two hemisphere catalogues, one based on the single-band 1B detection and the other based on the 3B detection method. During the merging, the overlap between tiles is removed. The catalogue retains only detections within the non-overlapping area of each tile, except for sources within a $\pm30\arcsec$ border region in overlapping areas. In these regions, source lists from adjacent tiles were cross-matched to prevent valid detections near tile boundaries from being discarded. A matching radius of $15\arcsec$ is used for point sources, and $30\arcsec$ for extended sources; this corresponds to roughly half and one HEW of the survey-averaged PSF, respectively.

The 1B source lists were generated with a low detection likelihood\footnote{The detection likelihood is defined as $L=-$\,ln\,$P$, where $P$ is the probability of the source being caused by random background fluctuation \citep[see][for details]{Brunner2022}.} threshold of {\tt DET\_LIKE\_0}~$\geq 5$ to maximise completeness. However, the main catalogue we release includes only point and extended sources with {\tt DET\_LIKE\_0}~$\geq 6$. Unlike \citet{Merloni2024}, we do not provide a supplementary list of sources with $5 \leq {\tt DET\_LIKE\_0} < 6$ given the high expected spurious detection rate. From the eRASS1 digital twin simulation\footnote{The digital twin simulation is a physically motivated, end-to-end numerical model that reproduces both the astrophysical sky and the instrumental response to create realistic synthetic observations of the eRASS1 survey.} \citep{Comparat2020}, the eRASS1 main catalogue is expected to contain about 14\% spurious sources \citep[Table 3 in][see also Sect.~\ref{sec:comparison_Xray_catalogues} below]{Seppi2022}. These results can also be used to estimate contamination levels in eRASS:3, since the detection likelihood in eSASS depends primarily on the count distribution rather than on exposure time. Consequently, sources at a fixed threshold (e.g. {\tt DET\_LIKE\_0}$~=6$) in eRASS:3 should exhibit similar contamination ($\sim14\%$) properties as in eRASS1, although the deeper data allow a larger number of sources to reach this threshold. 

For the hard catalogue, a more conservative threshold of {\tt DET\_LIKE\_3}~$\geq 12$ is adopted to limit the number of spurious detections in the $2.3-5.0$~keV band, where the higher background and lower sensitivity increase contamination \citep{Liu2022}. From \cite{Seppi2022}, we expect around $8-10$\% contamination by spurious sources in this hard band-selected sample.

Table~\ref{tab:detections} summarises the selection criteria and basic properties of the catalogues. Compared to eRASS1, the number of detected point-like sources ({\tt DET\_LIKE\_0~$\geq6$,~EXT\_LIKE~$=0$}) increases by more than a factor of two ($\approx 2.1$) in the eRASS:3 main source catalogue. In a similar way, the number of extended sources ({\tt DET\_LIKE\_0~$\geq6$,~EXT\_LIKE~$>0$}) increases by a factor of $\approx 2.4$ relative to eRASS1. The eRASS:3 hard source sample ({\tt DET\_LIKE\_3~$\geq12$,~EXT\_LIKE~$\geq0$}) is larger by a factor of $\approx 2.9$ compared to eRASS1. Section~\ref{sec:crossidentification_soft_hard} discusses in detail the comparison between eRASS1 and eRASS:3 catalogues. Figure~\ref{fig:net_counts_distribution} shows the distributions of net counts for the main and hard catalogues (top panels), as well as their cumulative version: distribution of the number of sources with net counts greater than or equal to a given value (bottom panels).

Any astrometric offsets remaining after calibrating the event positions with the time-dependent bore-sight calibration were removed by applying again the astrometric correction described in \ref{sec:tel_astrometry}. As expected, the remaining corrections in the ecliptical coordinate system $\lambda,~\beta$ are small, with mean values of $\Delta_{\lambda, \mathrm{mean}}\ = 0.03''$, $\Delta_{\beta, \mathrm{mean}}\ = 0.1''$ and standard deviations $\sigma_{\lambda} = 0.28'' $ and $\sigma_{\beta} = 0.16''$. 

After merging, 728 detections were classified as spurious based on their angular correlation with the remaining bad pixel events mentioned in Sect.~\ref{sec:known_issues}. 

The positional uncertainties of the eRASS:3 catalogues, encapsulated by the parameter {\tt RADEC\_ERR}, are tested using the Gaia/unWISE QSO sample as an external astrometric reference frame \citep{Shu2019}. The adopted methodology follows Sect.~6.2 of \cite{Merloni2024} and yields updated positional uncertainties for the X-ray sources, which are reported by the {\tt POS\_ERR} catalogue parameter defined as
\begin{equation}\label{eq:radial-dist.poserror_final}
    \mathrm{POS\_ERR} =  \sqrt{ 1.3 \cdot \sigma^2 + 0.4^2}.
\end{equation}
\noindent where $\sigma =\mathtt{RADEC\_ERR}/\sqrt{2}$ and {\tt RADEC\_ERR} is the catalogued positional uncertainty produced by \texttt{ermldet}. Appendix~\ref{appendix:catalogues} provides a summary of the content of the eRASS:3 main and hard catalogues, including descriptions of the columns and their associated units.

\begin{table}[]
    \caption{Basic eRASS:3 catalogue properties.}
     \renewcommand{\arraystretch}{1.2}
     \begin{tabular}{lccr}
    \hline
    \hline
    \multicolumn{4}{c}{1B detection [$0.2-2.3$ keV]} \\
    \hline
    Catalogue & {\tt DET\_LIKE\_0} & {\tt EXT\_LIKE} & \# of sources\\
    \hline
    Main, PS  & $\geqslant$6 & $=$0 & \num{1911744} \\
    Main, EXT  & $\geqslant$6  & $>$0 & \num{63796}  \\
    \hline
    \multicolumn{4}{c}{3B detection [$2.3-5.0$ keV only]} \\
    \hline
    Catalogue & {\tt DET\_LIKE\_3} & {\tt EXT\_LIKE} & \# of sources\\
    \hline
    Hard  & $\geqslant$12 &  $\geqslant$0 & \num{15980} \\
    Hard, PS & $\geqslant$12 &  $=$0 & \num{15026} \\
    Hard, EXT & $\geqslant$12 &  $>$0 & \num{954} \\
    \hline
    \end{tabular}
    \tablefoot{‘PS’ denotes point sources (\texttt{EXT\_LIKE} $=0$), while ‘EXT’ refers to extended sources with \texttt{EXT\_LIKE} $>0$. The definition of \texttt{EXT\_LIKE} differs slightly between the 1B and 3B detections: in 1B it is computed using photons in the $0.2-2.3$~keV band, whereas in 3B it is evaluated using all photons in the $0.2-5.0$~keV energy band. }
    \label{tab:detections}
\end{table}

\subsection{Flagging of problematic sources}
\label{sec:spurious_flagging}
The eRASS:3 catalogue contains spurious sources and inaccurate source properties due to limitations in the detection process. This also posed an issue in the eRASS1 catalogues, and to address it, \cite{Merloni2024} developed a flagging system to mark sources likely affected by high background noise or by misleading emission from bright or extended objects, such as supernova remnants, nearby galaxies, or star clusters. In the following, we briefly describe the flags that are included in the eRASS:3 main and hard catalogues. A summary of these flags is provided in Table~\ref{tab:flags}.

Overdense regions, where anomalously high source counts are observed, were empirically identified using flux thresholds and density maps, and then cross-matched with known astrophysical structures using the SIMBAD database. Four main flags are assigned for sources located in or near:
\begin{itemize}
    \item Supernova remnants ({\tt FLAG\_SP\_SNR});
    \item Bright point sources ({\tt FLAG\_SP\_BPS});
    \item Galactic star clusters ({\tt FLAG\_SP\_SCL});
    \item Local galaxies ({\tt FLAG\_SP\_LGA}).
\end{itemize}

Additional flagging targets sources near known galaxy clusters using the same list of published X-ray cluster catalogues described in \cite{Merloni2024}. A specific flag ({\tt FLAG\_SP\_GC\_CONS}) marks sources within a defined radius of these clusters, though optical and Sunyaev-Zel'dovich (SZ)-selected catalogues were excluded to avoid mismatches. Visual inspections and further cleaning steps were applied to the extended source catalogue to correct remaining misclassifications.

Table~\ref{tab:source_flag_summary} shows the number of sources for each of the flag categories in the eRASS:3 catalogues. After excluding flagged sources, the refined main catalogue contains $\num{1890579}$ point sources, and the hard catalogue contains $\num{14721}$. Users are still advised to manually inspect their sources of interest, especially in complex regions, before drawing scientific conclusions.

\subsection{Flagging of optical loading sources}
\label{sec:opt_flagging}
The eROSITA detectors are prone to optical loading. In this process, low-energy photons are accumulated within a CCD pixel over the integration time of 50~ms and if the summed energy of optical photons plus detector noise exceeds the X-ray detection threshold, a `false' X-ray event is generated. Likewise, optical photons might add energy or distort the pattern of real X-ray photons. The relevance of these, in detail quite complex, effects is mainly determined by the brightness of the respective stellar source. The detector position also matters significantly, but is mostly averaged out in survey observations. A detailed discussion of optical loading and its effects is presented in Robrade et al. (in prep.) and \cite{Zheng2026}. To address this issue, we cross-match the eRASS:3 catalogues with optical sources, following a similar approach to that for eRASS1 \citep{Merloni2024}.

The {\tt FLAG\_OPT} flag, as included in the eRASS:3 catalogues, is set if an eRASS X-ray source is located within a distance of $15\arcsec$ to an optically bright source. In the cross-matching we used the third Gaia data release \citep[GDR3,][]{gdr3}, Tycho-2 \citep{tycho2}, 2MASS \citep{2mass}, plus the Simbad database. Updated epoch\,=\,J2021 positions were used as input when proper motion is available. The brightness limits in the photometric bands are adjusted to accommodate the deeper exposure of eRASS:3 and are half a magnitude fainter than those for eRASS1. Specifically, $B, V, G \le 5.0$~mag and $J \le 3.5$~mag are used. If one or more of these criteria are met, the source is flagged; a total of $1\,176$ such sources were selected. Note that the presence of the optical flag does not necessarily indicate that the specific X-ray source is affected by optical loading, but additional caution is advised.

\begin{table*}[htbp]
    \centering
    \caption{Spurious and problematic source flags description.}\label{tab:spurious_source_flag_descriptions}
    \renewcommand{\arraystretch}{1.2}
    \begin{tabular}{p{3cm}|p{13cm}}
        \hline
        \hline
        Flag name & Description \\
        \hline
         \texttt{FLAG\_SP\_SNR} & Source might lie within an overdense region near a supernova remnant.\\
         \texttt{FLAG\_SP\_BPS} & Source might lie within an overdense region near a bright point source. \\
         \texttt{FLAG\_SP\_SCL} & Source might lie within an overdense region near a stellar cluster.\\
         \texttt{FLAG\_SP\_LGA} & Source might lie within an overdense region near a local large galaxy.\\
         \texttt{FLAG\_SP\_GC\_CONS} & Source might lie within an overdense region near a galaxy cluster.  \\
         \texttt{FLAG\_NO\_RADEC\_ERR} & Source contained no \texttt{RA\_DEC\_ERR} in the pre-processed version of the eSASS catalogue. \\
         \texttt{FLAG\_NO\_CTS\_ERR} & Source contained no \texttt{ML\_CTS\_ERR\_1} in the pre-processed version of the eSASS catalogue. \\
         \texttt{FLAG\_NO\_EXT\_ERR} & Source contained no \texttt{EXT\_ERR} in the pre-processed version of the eSASS catalogue. \\
    \hline
    \end{tabular}
    \label{tab:flags}
\end{table*}

\begin{figure*}
    \centering
    \includegraphics[width=0.4\textwidth]{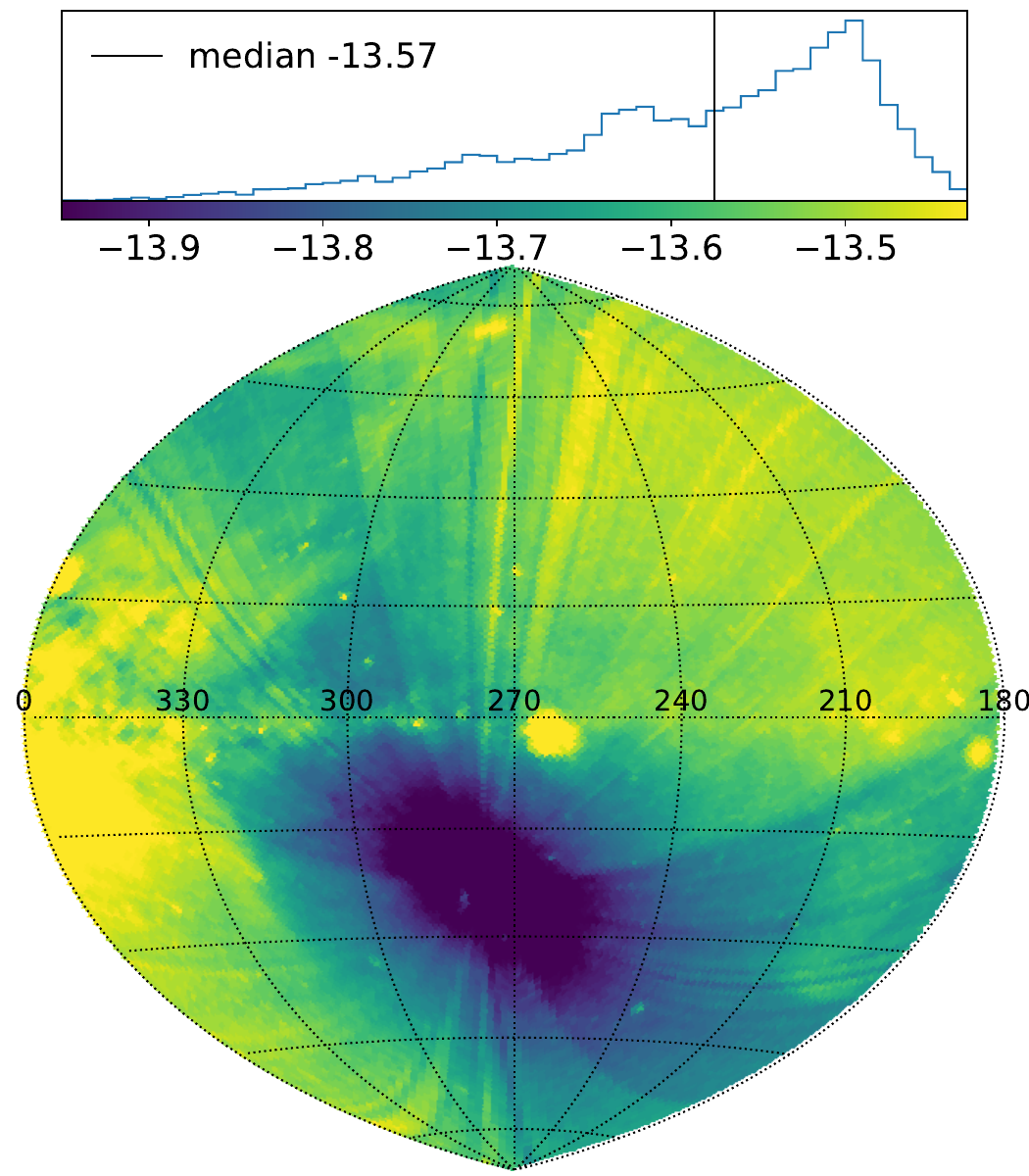}
    \hspace{1cm}
    \includegraphics[width=0.4\textwidth]{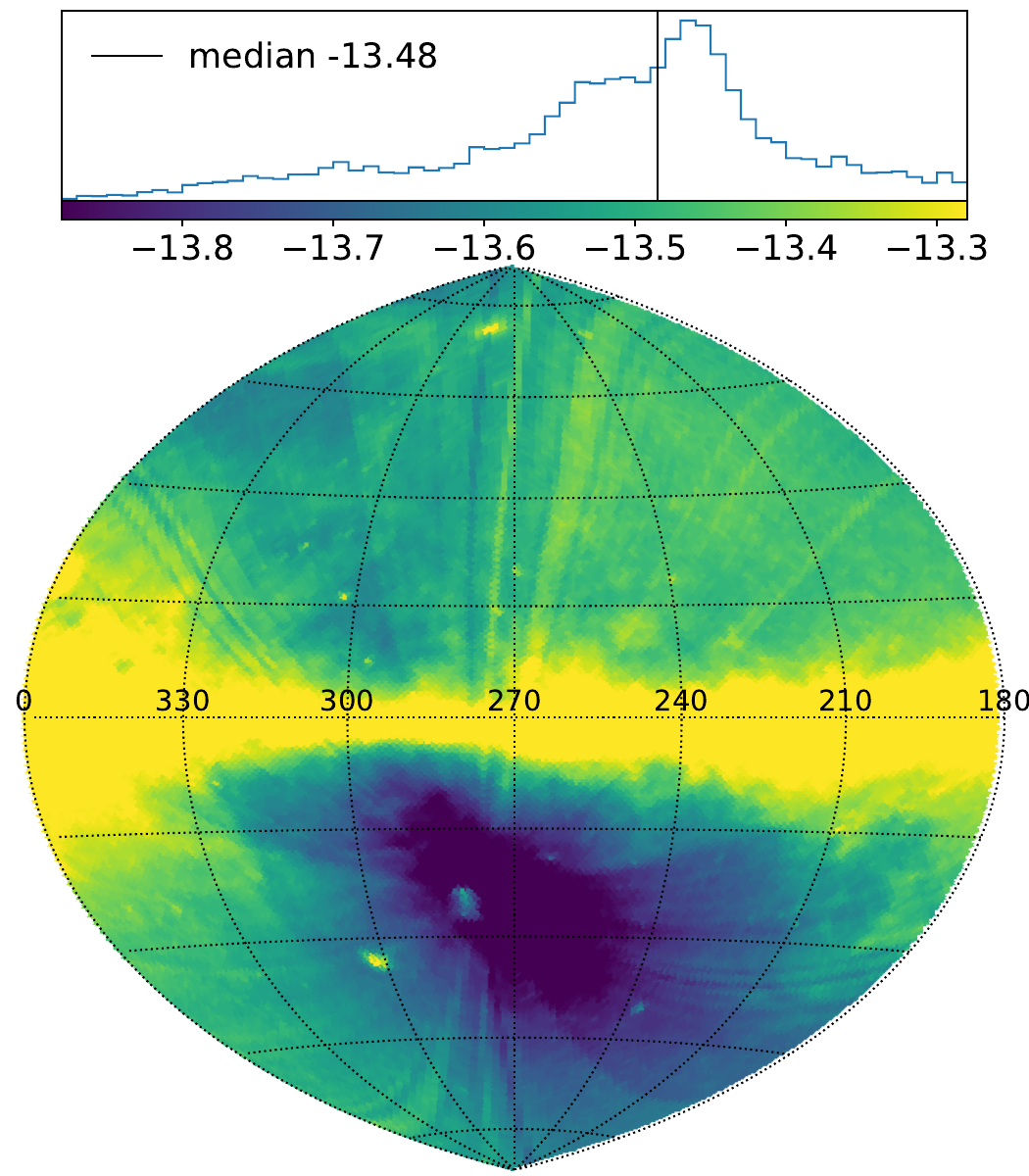}
    \caption{Hammer–Aitoff projection maps in Galactic coordinates showing the logarithm of the $0.5-2.0$~keV flux limit, defined as the flux corresponding to a 50\% sky covering fraction. The left and right panels display the values uncorrected and corrected for Galactic absorption, respectively. The histograms above each map show the distribution of $\log(F_{\mathrm{lim}})$ in units of erg\,s$^{-1}$\,cm$^{-2}$. The colour bar along the X-axis indicates the intensity scale corresponding to the flux limits in the projected maps.}
    \label{fig:flux_limit_soft_band}
\end{figure*}

\subsection{Cross-identification of soft and hard band sources}
\label{sec:crossidentification_soft_hard}
As described in Sect.~\ref{subsec:det_chain}, source detection was performed in two configurations: a single-band (1B, $0.2-2.3$~keV energy band) and a three-band setup (3B, $0.2-0.6$, $0.6-2.3$, and $2.3-5.0$~keV energy bands), with the latter subsequently filtered based on significance in the $2.3-5.0$~keV band. These yield soft- and hard-band–selected source samples, respectively. The datasets from the 1B and 3B detections largely overlap; the only additional photons in the 3B case are from the $2.3-5.0$~keV range, where the instrumental effective area is lower, and the background is higher. Consequently, most sources in the hard catalogue have corresponding counterparts in the main catalogue. 

In eFEDS, \cite{Nandra2025} showed that $\sim 90\%$ of the sources in the hard catalogue have a counterpart in the main catalogue. Similarly, \cite{Merloni2024} reported that $\sim 84\%$ of the hard point-like sources in eRASS1 have an associated soft-band detected source. We employed the same matching algorithm between the eRASS:3 main and hard catalogues as described in \cite{Merloni2024}. The algorithm results in weak and strong source matches. A weak match requires a maximum separation of $10\arcsec$ for point sources or 4 times the extent value for extended sources, and also allows matches between point and extended sources. Strong matches require the sources to be either point-like or extended in both catalogues. The point source matching radius is set to 3 times the larger {\tt RADEC\_ERR} value of the pair, with a maximum radius of $10\arcsec$. In addition, for point sources, the $0.2-2.3$ keV count values are not allowed to disagree by more than $50\%$ for a strong match. For a strong match between extended sources, they must be detected as extended in both catalogues, and the separation must be less than $4\times$ the \texttt{EXT} value in the hard catalogue. We record the unique identifier (UID) of associated counterparts in the {\tt UID\_Hard} column of the main catalogue and in the {\tt UID\_1B} column of the hard catalogue to flag strongly associated sources. Weak associations are encoded by storing the corresponding UID multiplied by $-1$ in the same columns. Consequently, positive values of {\tt UID\_Hard} or {\tt UID\_1B} indicate strong associations, while negative values denote weak (possible) counterparts in the other catalogue. A value of {\tt UID\_1B}$\ =0$ identifies hard sources with no counterpart within the matching radius in the 1B catalogue.

Table \ref{tab:main_hard_matches} summarises the results of the matching between the eRASS:3 main and hard catalogues. In the hard catalogue, 1\,299 point sources have no match in the main catalogue and can be considered candidates for sources with a very hard X-ray spectrum. However, only 104 of these hard-only detections have matching counterparts in the eRASS1 hard catalogue \citep[eRASS1\_hard,][]{Merloni2024}. Even when considering that the eRASS1 hard catalogue is shallower, this clearly indicates that the selection of hard-only detections results in a very high fraction of spurious detections. This effect is also discussed by \cite{Waddell2026} in the context of the eRASS1 hard catalogue.

\begin{table}[]
    \centering
    \caption{eRASS:3 hard matches to eRASS:3 main sources.}
     \renewcommand{\arraystretch}{1.2}
     \begin{tabular}{lrr}
        \hline
        \hline
        Match type & PS & EXT \\
        \hline
        Weak   &  \num{112} &  \num{30} \\
        Strong &  \num{13615} &  \num{892}\\
        None   &   1\,299 &  \num{32} \\      
        \hline
    \end{tabular}
    \tablefoot{‘PS’ and ‘EXT' denote eRASS:3 point sources and extended sources from the hard catalogue matching with detections from the eRASS:3 main catalogue.}
    \label{tab:main_hard_matches}
\end{table}

\subsection{The flux limit of eRASS:3 in the $0.5-2.0$~keV energy band}
\label{sec:flux_limit}
We determined the eRASS:3 flux limit in the $0.5-2.0$~keV energy band following the method described by \citet{Merloni2024}. A brief summary of this procedure is provided below.

The local flux limit is defined from the sensitivity curve as the flux at which the probability of detecting a source at the adopted threshold reaches $50\%$. To derive this curve, aperture photometry measurements at source positions obtained using the eSASS task {\tt apetool} are employed. Energy conversion factors (ECFs) are then used to convert count rates into fluxes, assuming a power-law spectrum with photon index $\Gamma=2$ and position-dependent Galactic hydrogen column densities from \cite{Willingale2013}. Forced photometry in the $0.5-1.0$~keV (band P2; see Appendix~\ref{appendix:catalogues} for the band definitions) and $1.0-2.0$~keV (band P3) energy bands is combined to obtain soft-band quantities for both sources and background regions. The total exposure in the source and background apertures is computed by weighting and combining the vignetted exposure in P2 and P3. Using these measurements, the source and background count rates, together with the ECFs and Poisson detection probabilities, are computed using the \texttt{scipy.special}\footnote{Python package: \url{https://docs.scipy.org/doc/scipy/reference/special.html}} package. Two sets of these quantities were computed: corrected and uncorrected for Galactic absorption.

The photometry is performed using apertures that enclose $75\%$ of the PSF and adopting a Poisson false-detection probability threshold of $P_{\mathrm{th}}=4\times10^{-6}$ \citep{Georgakakis2008}. Flux-limit maps are produced by dividing the sky into HEALPix\footnote{\url{https://healpix.sourceforge.io/}} pixels (order 6) and tessellating sources via Voronoi cells, with the average of local sensitivity curves within each pixel weighted by the corresponding cell area. The flux corresponding to a $50\%$ detection probability from the mean sensitivity curve is finally adopted as the local limit.

In our work, two distinct flux-limit maps were produced for eRASS:3, as illustrated in Fig.~\ref{fig:flux_limit_soft_band}. The left panel shows the flux limit for X-ray point sources without accounting for Galactic absorption, while the right panel corresponds to the absorption-corrected case. The median flux limits across the hemisphere are approximately $2.7\times 10^{-14}$~erg s$^{-1}$ cm$^{-2}$ (uncorrected) and $3.3\times 10^{-14}$~erg s$^{-1}$ cm$^{-2}$ (absorption-corrected). The eRASS:3 survey achieves a flux limit approximately two times deeper than that of eRASS1 \citep{Merloni2024}, corresponding to roughly half the limiting flux reported for the first all-sky survey. eROSITA-DE DR2 makes available eRASS:3 flux-limit information to users via the upper-limit server described in Sect.~\ref{sec:dr2_description}.

\subsection{log\,N--log\,S distribution of point sources}
\label{sec:logN_logS}
The X-ray point-source number density exhibits spatial variations across the sky in eRASS1 \citep{Merloni2024}. These variations arise primarily from inhomogeneous Galactic absorption and the non-uniform distribution of Galactic X-ray sources, with additional contributions from large-scale structure and possible anisotropies in the extragalactic AGN population. In the eRASS:3 survey, we therefore examine the distributions of point-source counts, averaged over broad Galactic latitude ranges.

\begin{figure}
    \centering
    \includegraphics[width=0.48\textwidth]{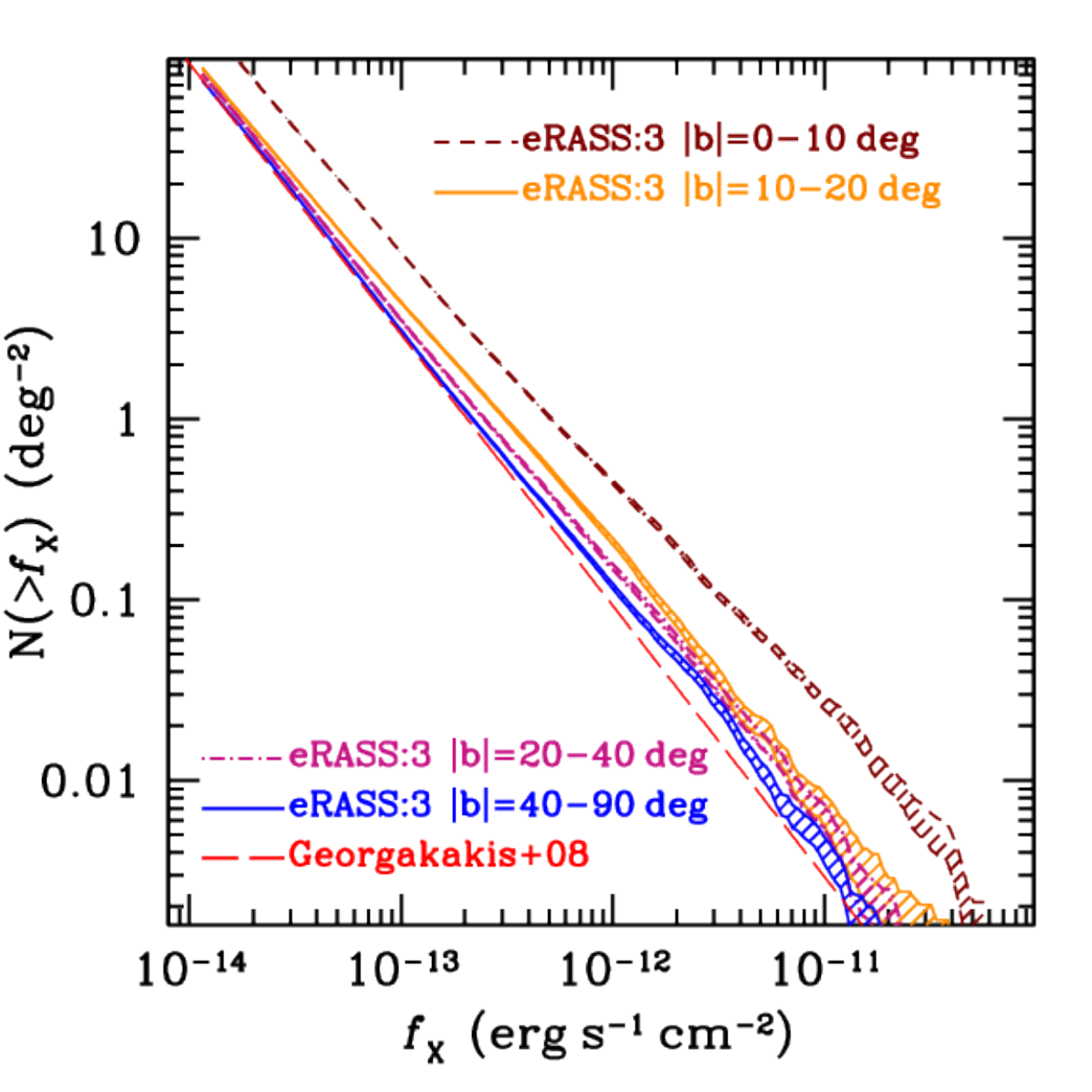}
    \caption{Cumulative number counts as a function of flux for eRASS:3 X-ray point sources in the $0.5-2.0$~keV energy band. The number of count distributions is shown for different Galactic latitude intervals: $|b|=0-10^{\circ}$ (brown dashed curves), $10-20^{\circ}$ 
(amber solid curves), $20-40^{\circ}$ 
(purple dashed curves), and $40-90^{\circ}$ 
(blue dashed curves). Shaded regions indicate the $68\%$ uncertainties estimated via bootstrap resampling. The $0.5-2.0$~keV number counts from \citet{Georgakakis2008} are shown as a dashed red line for comparison.}
    \label{fig:logNlogS}
\end{figure}

We divided the hemisphere into four Galactic latitude bins: $0-10^{\circ}$, $10-20^{\circ}$, $20-40^{\circ}$, and $40-90^{\circ}$. Point-source number counts were computed following \cite{Georgakakis2008} and \cite{Merloni2024} using the {\tt apetool} products of the 1B catalogue in the $0.2-2.3$~keV energy band, converting count rates to absorption-corrected fluxes via the $N_{\rm H}$-dependent ECF. Uncertainties are estimated through bootstrap resampling, generating 100 realisations per subsample and deriving the $1\sigma$ scatter at fixed flux. The resulting cumulative number counts in the $0.5-2.0$~keV energy band are shown in Fig.~\ref{fig:logNlogS}. These curves include a systematic shift of $\log_{10}(0.57)$, in the faint flux direction, to convert the flux in the energy band $0.2-2.3$\,keV to $0.5-2.0$\, keV and thereby enable comparisons with literature results \citep[e.g.][]{Georgakakis2008}. The conversion factor above is estimated assuming a power-law spectral energy distribution with spectral index $\Gamma=1.9$. The plotted number counts are in good agreement with those presented in \cite{Merloni2024} as part of eRASS DR1.

%
%

\begin{table*}[t]
  \centering
  \caption{Matching eRASS:3 main detections with other X-ray catalogues.}
  \renewcommand{\arraystretch}{1.2}
  \begin{tabular}{lrrrrrrrr}
    \hline
    \hline
       &  \multicolumn{2}{c}{eRASS1} &  \multicolumn{2}{c}{2RXS}  &  \multicolumn{2}{c}{5XMM-DR15}  &  \multicolumn{2}{c}{CSC 2.1} \\
       Match type & \multicolumn{1}{c}{PS} & \multicolumn{1}{c}{EXT} & \multicolumn{1}{c}{PS} & \multicolumn{1}{c}{EXT} & \multicolumn{1}{c}{PS} & \multicolumn{1}{c}{EXT} & \multicolumn{1}{c}{PS} & \multicolumn{1}{c}{EXT} \\
    \hline
        Strong     &  \num{689050} &  \num{19257} &  \num{31824}   &   2\,022 &  \num{51081} & 6\,265 &  \num{17940} & 1\,751  \\
        Weak       &   \num{10324} &  \num{23425} &  5\,431 &   4\,993  &  4\,114 & 2\,510  &   2\,458 & 1\,685\\
        None        & \num{1212370} & \num{21114} &  \num{1874489} & \num{56781}  &   \num{20122} & \num{698} &    4\,256 & 54 \\
        Not covered & -       & -     & -  &  -           &  \num{1836427}& \num{54323} &  \num{1887090} & \num{60306} \\
    \hline
  \end{tabular}
  \tablefoot{‘PS’ and ‘EXT' denote eRASS:3 point-like sources and extended sources, respectively. }
  \label{tab:external_matches}
\end{table*}

\section{Comparison with other catalogues}
\label{sec:comparison_Xray_catalogues}
The algorithm applied to match the eRASS:3 main and hard catalogues (see Sect.~\ref{sec:crossidentification_soft_hard}) was also applied to match the eRASS:3 main catalogue with: the eRASS1 main catalogue \citep{Merloni2024}, the second ROSAT all-sky survey catalogue \citep[2RXS][]{Boller2016}, the XMM-{\it Newton} 5XMM-DR15 catalogue (Webb et al., in prep.), and the {\it Chandra} CSC 2.1 catalogue \citep{Evans2024}. The parameters defining the criteria for weak and strong matching were adapted to the properties of the comparison catalogues. Since these catalogues are based on observations spanning a wide range of epochs, no flux criterion was applied for matching.

Criteria for weak and strong association were defined in a similar way as described in Sect.~\ref{sec:crossidentification_soft_hard}. For eRASS1 and XMM-{\it Newton} 5XMM-DR15\footnote{We cross-matched the eRASS:3 catalogue with 4XMM-DR14 \citep{Webb2020} using the same association algorithm as with 5XMM-DR15. The resulting matching statistics are highly consistent, with minor differences attributable to the 4XMM's more limited footprint.}, the matching radii for weak associations were $16\arcsec$ for point sources, or 4 times the larger of the two extent values for pairs with at least one extended source. For strong point-source associations, matching radii of $4$ times the combined positional errors, ranging from $5\arcsec$ to $16\arcsec$, were used. The matching radius for strong associations between two extended sources is $4$ times the larger of the two extent values. The same radii were used for matching with the CSC 2.1 catalogue, where the positional errors and source extent values were derived by averaging the 2-dimensional \texttt{err\_ellipse\_r0, err\_ellipse\_r1} (corrected to $ 1\sigma$) and \texttt{minor\_axis\_b, major\_axis\_b} values. The 2RXS catalogue does not contain positional error estimates. For the matching with this catalogue, the matching radii are $40\arcsec$ (point-like) or $4$ times the extent (extended) for weak associations and $30\arcsec$ (point-like) and max($60$~arcec,~4$\times$ \texttt{EXT}) (extended) for strong associations.

The source identifiers from the matching catalogue entries are stored in the following columns of the eROSITA DR2 catalogue: \texttt{UID\_DR1} (eRASS1 DR1 \texttt{UID}), \texttt{UID\_2RXS} (2RXS \texttt{IND\_2RXS}), \texttt{UID\_5XMM} (5XMM \texttt{SRCID)}, and \texttt{UID\_CSC} (CSC \texttt{name}). These columns contain a catalogue identifier ($>0$) for strong associations, a negative value (identifier multiplied by $-1$) for weak associations or $0$ for no association. The column \texttt{UID\_5XMM} contains $-1$ in case the position of the eROSITA source is outside the footprint of the 5XMM-DR15 catalogue. For the CSC 2.1, where the source identifier is a string containing the IAU source designation, the column \texttt{FLAG\_CSC} was added. This column contains the value $1$ for a strong association, $-2$ for a weak association, 0 for no association in the CSC footprint, and $-1$ for eRASS:3 sources outside the CSC 2.1 footprint. Table \ref{tab:external_matches} shows the distribution of matching categories in the eRASS:3 main catalogue for point-like and extended detections.

The same methods were used to correlate the eRASS:3 hard catalogue with the eRASS1 hard (DR1), 2RXS, 5XMM-DR15, and CSC catalogues. The results are summarised in Table \ref{tab:external_matches_hard}. 

\begin{figure}
    \centering
    \includegraphics[width=0.5\textwidth,trim={0 0 -10 0},clip]{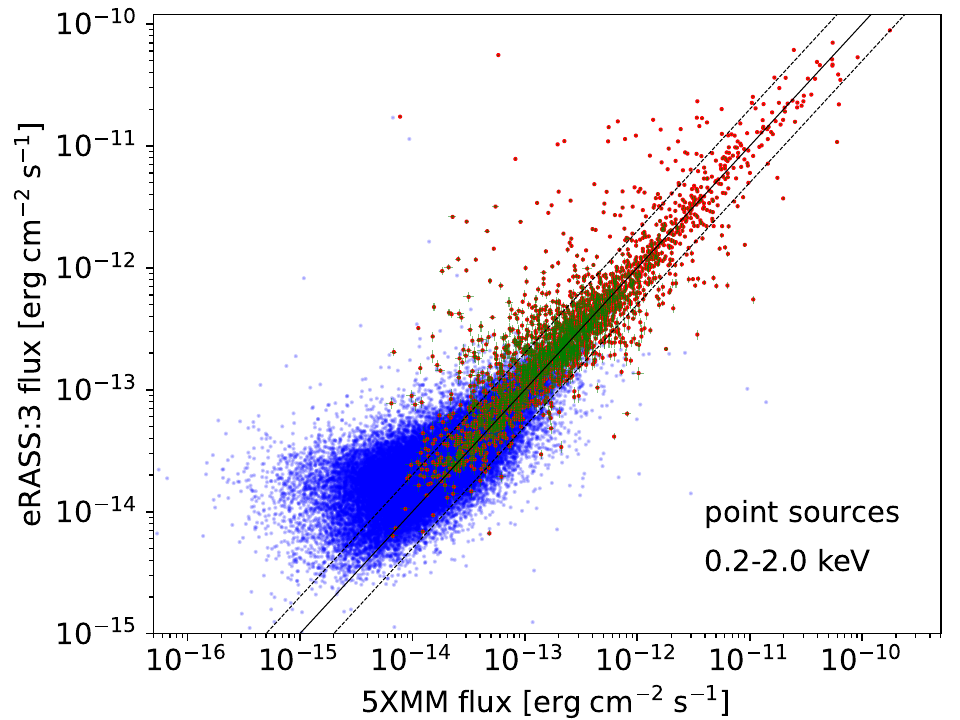}
    \caption{Comparison between eRASS:3 main catalogue point-source fluxes and 5XMM-DR15 fluxes (both in the $0.2-2.0$~keV band) for sources with strong matches between the two catalogues. Blue dots represent the full sample of matched sources, while red dots (with green error bars) mark sources with $8 \sigma$ flux detections in both catalogues. The solid line indicates the one-to-one relation, and the dashed lines correspond to flux ratios of 2:1 and 1:2, respectively.}
    \label{fig:fluxes_5xmm_pnt}
\end{figure}

\begin{figure}
    \centering
     \includegraphics[width=0.5\textwidth,trim={0 0 -10 0},clip]{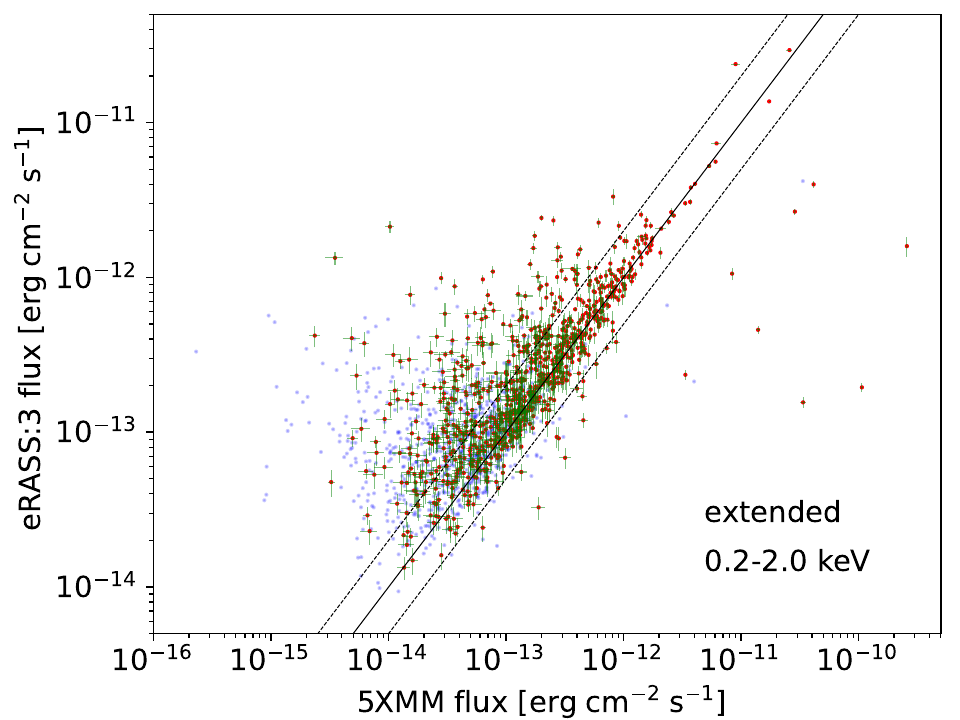}
    \caption{Same as Fig.~\ref{fig:fluxes_5xmm_pnt} but for extended sources. Red dots (with green error bars) mark sources with $5 \sigma$ flux detections in both catalogues.}
    \label{fig:fluxes_5xmm_ext}
\end{figure}

\begin{figure}
    \centering
    \includegraphics[width=0.5\textwidth,trim={0 0 -10 0},clip]{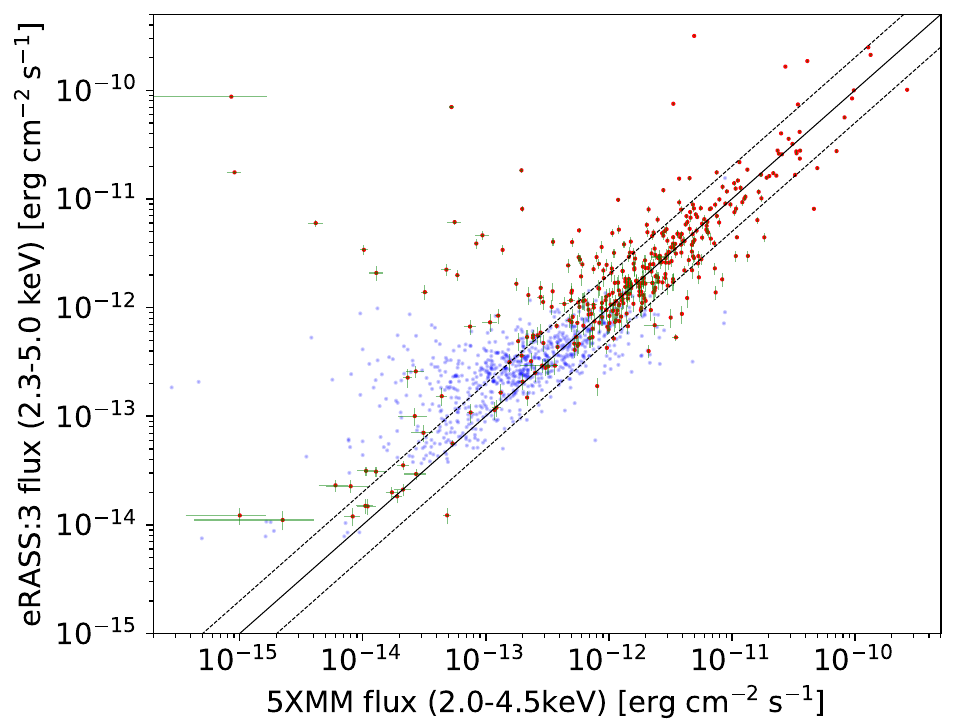}
    \caption{Comparison between eRASS:3 hard catalogue point source fluxes ($2.3-5.0$~keV) and 5XMM-DR15 fluxes ($2.0-4.5$~keV) for sources with strong matches between the two catalogues. Blue dots represent the full sample of matched sources, while red dots (with green error bars) mark sources with $5 \sigma$ flux detections in both catalogues. The solid line indicates the one-to-one relation, and the dashed lines correspond to flux ratios of 2:1 and 1:2, respectively.}
    \label{fig:fluxes_5xmm_hard_pnt}
\end{figure}

Figures \ref{fig:fluxes_5xmm_pnt} - \ref{fig:fluxes_5xmm_hard_pnt} show the comparisons of eRASS:3 source fluxes with respect to their 5XMM-DR15 counterparts for the main point source and extended source samples, and for the hard point source samples. For the bulk of the sources, the eROSITA fluxes are consistent with the 5XMM-DR15 fluxes. Strong flux deviations are typically due to source variability for point sources. Discrepant fluxes between matched pairs of extended detection are often due to detection issues, such as fragmented detection of large extended sources.

\begin{table*}[t]
  \centering
  \caption{Matching eRASS:3 hard detections with other catalogues.}
  \renewcommand{\arraystretch}{1.2}
  \begin{tabular}{lrrrrrrrr}
    \hline
    \hline
       &  \multicolumn{2}{c}{eRASS1hard} &  \multicolumn{2}{c}{2RXS}  &  \multicolumn{2}{c}{5XMM-DR15}  &  \multicolumn{2}{c}{CSC 2.1} \\
       Match type & \multicolumn{1}{c}{PS} & \multicolumn{1}{c}{EXT} & \multicolumn{1}{c}{PS} & \multicolumn{1}{c}{EXT} & \multicolumn{1}{c}{PS} & \multicolumn{1}{c}{EXT} & \multicolumn{1}{c}{PS} & \multicolumn{1}{c}{EXT} \\
    \hline
        Strong     & 4\,229 &  369  &   4\,597 &   333 &   1269 &  384 &     457 & 267 \\
        Weak       &   57 &  30   &    809 &   299  &   115 &   15 &     319 &  24 \\
        None       & \num{10740} & 555 &    9\,620 &   322  &    47 &    4 &      34 &   2 \\
        Not covered & --    & --   & --       &  --     & \num{13595} &  551 &   \num{14216} & 661 \\
    \hline
  \end{tabular}
  \tablefoot{‘PS’ and ‘EXT' denote eRASS:3 (hard) point-like sources and extended sources, respectively. The eRASS1 hard UIDs are stored in column \texttt{UID\_DR1Hard}.}
  \label{tab:external_matches_hard}
\end{table*}

\subsection{An independent assessment of spurious source contamination in eRASS1}
\label{sec:erass1_comparison}
The case of the comparison with the eRASS1 catalogue differs from the others and deserves special attention. In fact, the data used to generate the two catalogues are not independent; rather, we can assume that the data from which the eRASS1 catalogues were generated are fully contained in the dataset used to generate the eRASS:3 catalogues (modulo small differences introduced by the new 030 pipeline processing version). Thus, the comparison with the eRASS:3 catalogue represents a unique opportunity to assess the quality of the eRASS1 one, and in particular to provide a direct estimate of its contamination level, independent of the simulations of \citet{Seppi2022}.

\begin{figure*}
\sidecaption
    \includegraphics[width=5.92cm]{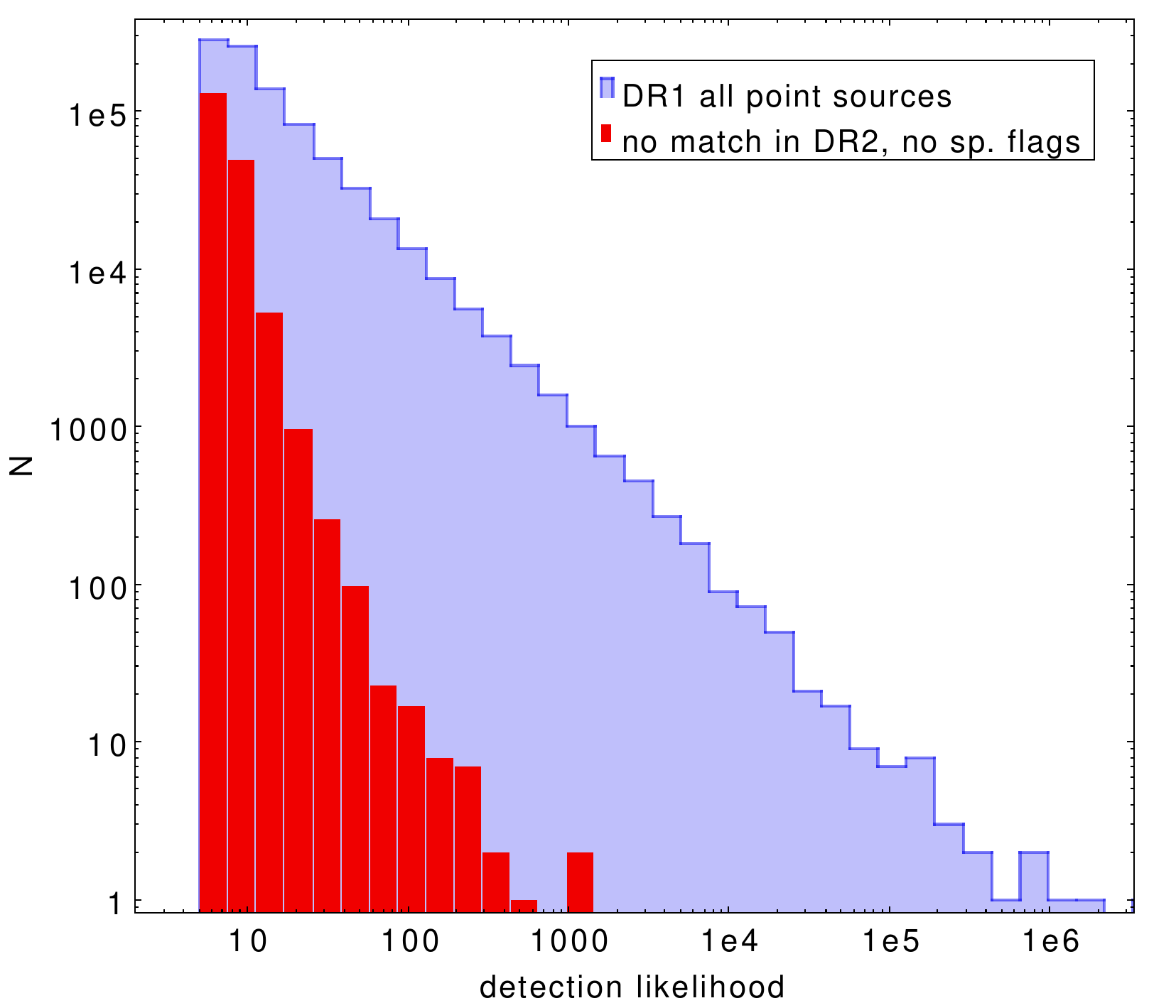}
    \includegraphics[width=5.8cm]{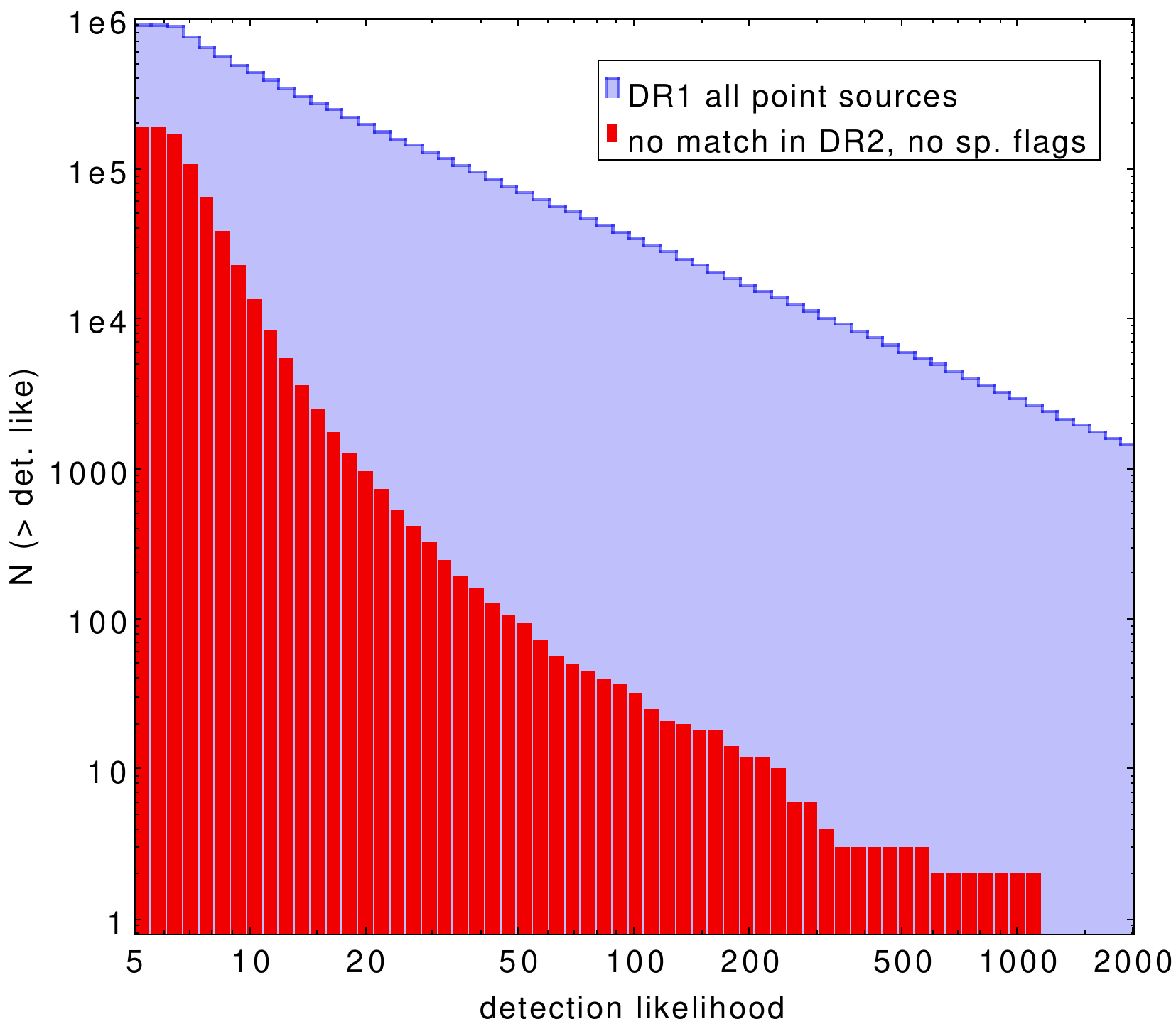}
    \caption{Detection likelihood distribution for the entire eRASS1 point source catalogue (blue histograms) and for the sources with no spurious flags and no match in eRASS:3 (red histograms). {\it Left panel:} Differential distribution; {\it Right panel:} Cumulative distribution. While in the entire catalogue the fraction of point sources not matched to eRASS:3 is about 21\%, this drops to about 3.5\% for \texttt{DET\_LIKE}$>$10 and 0.15\% for \texttt{DET\_LIKE}$>$50.}
    \label{fig:erass1_nomatch_histo}
\end{figure*}

\begin{figure*}
\sidecaption
    \includegraphics[width=5.9cm]{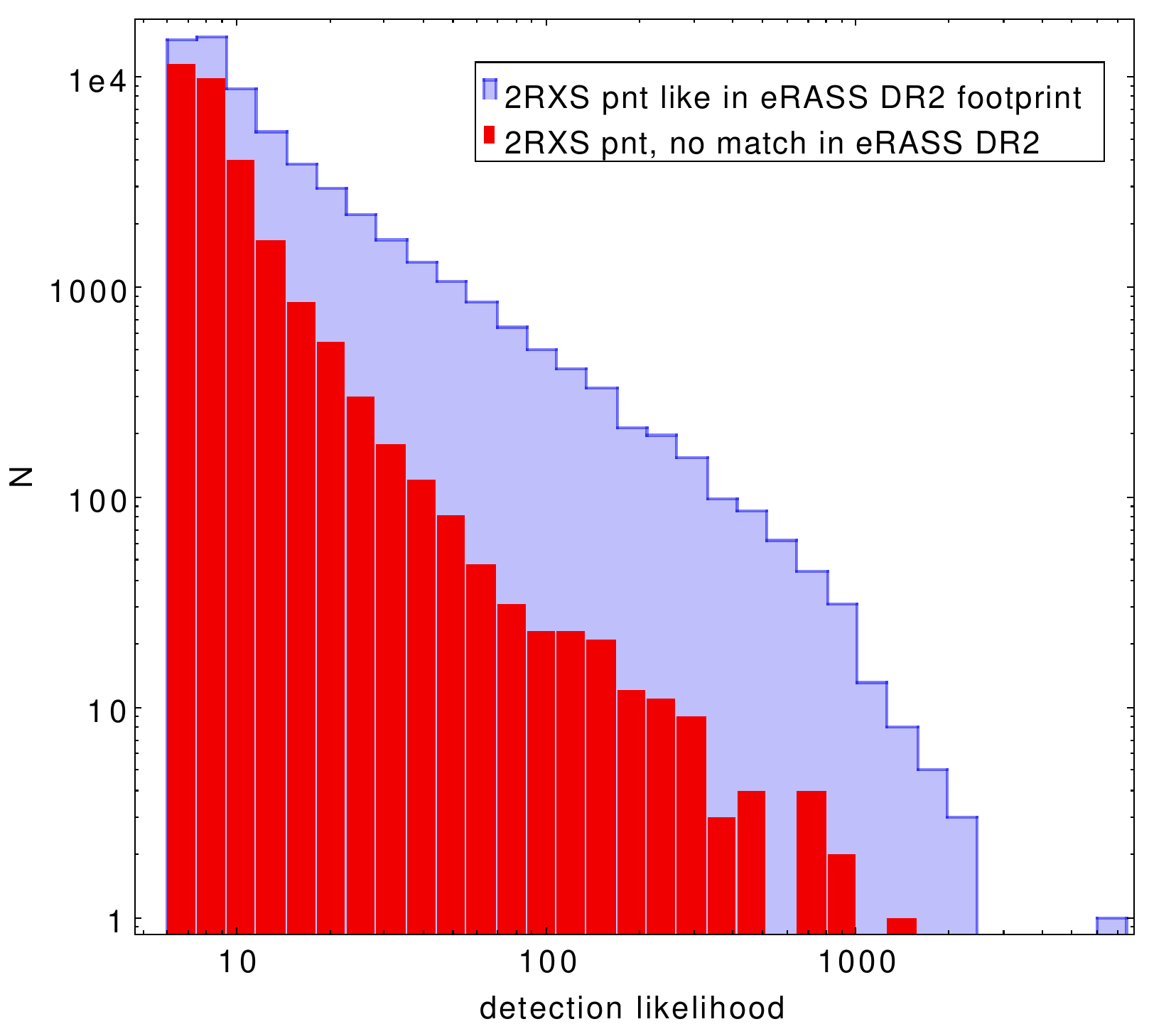}
    \includegraphics[width=5.9cm]{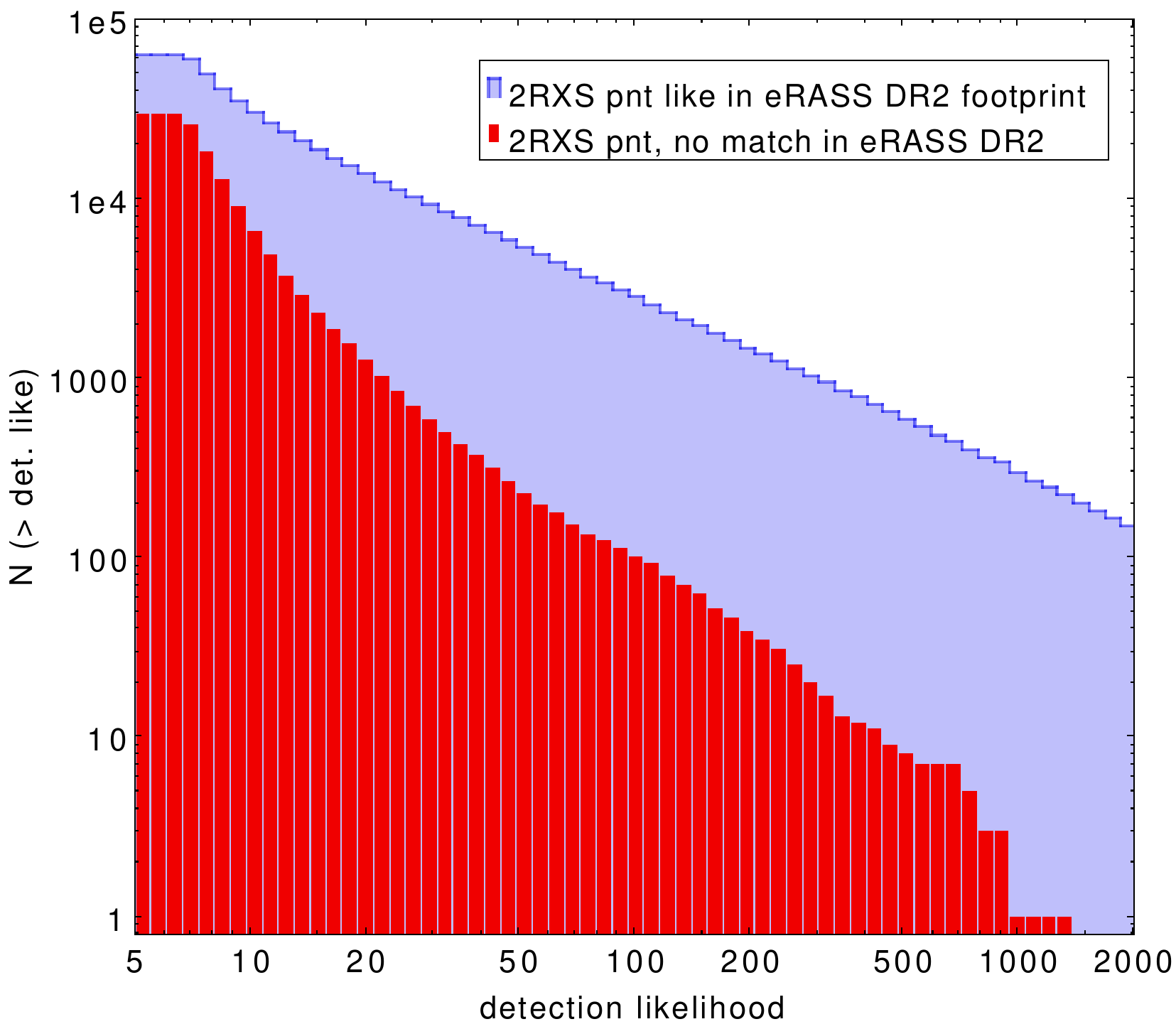}
    \caption{Detection likelihood distribution for the ROSAT 2RXS source catalogue over the eRASS DR2 footprint (Western Galactic Hemisphere, blue histograms) and for the sources with no spurious flags and no match in eRASS:3 (red histograms). {\it Left panel:} Differential distribution; {\it Right panel:} Cumulative distribution.}
    \label{fig:rosat_nomatch_histo}
\end{figure*}

From Table~\ref{tab:external_matches}, we see that only \num{699374} eRASS1 `main' sources are matched to eRASS:3, which is about 76\% of the total. If we further consider only point-like sources in eRASS1, and discard all those flagged as potentially spurious, we find that \num{187050}/\num{890036} (21\%) do not have a match in eRASS:3. This fraction is significantly larger than expected from the simulation work of \citet{Seppi2022}, which predicts instead $\sim14$\%. Figure~\ref{fig:erass1_nomatch_histo} shows the distribution of the detection likelihood of all DR1 point sources and of those without a match in eRASS:3, both differentially (left panel) and cumulative (right panel). Clearly, the sources without a match eRASS:3 are concentrated at low detection likelihood, but this does not imply that they are all spurious. Indeed, due to the Poisson nature of the X-ray detection process and the intrinsic variability of X-ray sources, it is possible that an object that barely exceeds our detection threshold in one survey might fall below threshold when multiple surveys are combined. 

We can assess the above by looking at the multi-wavelength counterparts of the eRASS1 sources. As shown in \citet[][see also Sect.~\ref{subsec:ctp} below]{Salvato2025}, the most complete identification of the eRASS1 X-ray sources' counterparts is available over the Legacy Survey DR10 \citep[LS10;] []{Dey2019} extragalactic area. Examining the LS10 counterparts of the \num{187050} non-matched sources, we find a clear bimodality in their \texttt{p\_any} distribution with 60\% of them having a very low value of \texttt{p\_any} (below 0.1), indicating they do not have a secure counterpart in the deep LS10 imaging catalogues. This \texttt{p\_any} threshold corresponds approximately to a purity \texttt{NWAY\_purity6}\,$>0.9$ \citep[see][and section~\ref{subsec:ctp}]{Salvato2025}.
Under the simplifying, but realistic assumption that these low-\texttt{p\_any} sources are all spurious detections, we conclude that the spurious contamination level of the eRASS1 point source catalogue is about 12.6\%, in very good agreement with the simulations-based estimate of \citet{Seppi2022}. On the other hand, the remaining $\sim$8.4\% of the eRASS1 sources without a match in eRASS:3, detected with low likelihood only in the first eROSITA all-sky survey, are likely genuine X-ray sources affected by intrinsic or Poisson-induced variability. In this way, we have shown how the combination of the cross-match with eRASS:3 and the multi-wavelength probabilistic associations of \citet{Salvato2025} combined can be used to further refine the quality of the published eRASS1 catalogue.

\subsection{About ROSAT/2RXS sources not matched to eRASS:3}
\label{sec:rosat_comparison}
Repeating the same experiment above with the ROSAT 2RXS catalogue (limited to the Western Galactic Hemisphere only; Fig.~\ref{fig:rosat_nomatch_histo}), we see that a higher fraction of 2RXS sources (\num{29148}/\num{60856}, 47.9\%) do not have a match in eRASS:3. Differently from the case of eRASS1, 2RXS sources are based on a dataset which is completely independent from the eRASS:3 one, and therefore source variability must play a significant role. Indeed, for detection likelihood larger than 20, almost 10\% of the 2RXS sources do not have a counterpart in eRASS:3. This is again broadly consistent with the expectation that the full 2RXS catalogue, down to the lowest detection likelihood, is contaminated by a fraction of spurious sources as high as $30-40$\% \citep{Boller2016}.

\subsection{Known issues and recommendations for catalogue users}
\label{sec:cat_issues}
 
In this section, we summarise the known issues in the eROSITA DR2 catalogues and provide recommendations for filtering and cleaning them. The spurious-source fractions estimated for the DR1 catalogue, despite its shallower depth, can be used as a guideline when choosing a detection-likelihood threshold (\texttt{DET\_LIKE\_0}) for the DR2 main catalogue (see Sect.~\ref{sec:erass1_comparison}).

\begin{table}
    \centering
    \caption{Source flags show the fractions of sources (\texttt{DET\_LIKE\_0} > 10) with strong 5XMM-DR15 associations within the 5XMM footprint.}\label{tab:spurious_source_flag_assoc}
    \renewcommand{\arraystretch}{1.2}
    \begin{tabular}{l|c|c}
        \hline
        \hline
        Flag name & \multicolumn{2}{l}{\% strong 5XMM associations} \\
                &  Point-like & Extended \\                
        \hline
         All in XMM FOV         & 82.6 & 71.2 \\
         \texttt{FLAG\_SP\_SNR} & 30.8  & 86.5 \\
         \texttt{FLAG\_SP\_BPS} &  43.8 & 53.9 \\
         \texttt{FLAG\_SP\_SCL} &  90.0 & 77.4 \\
         \texttt{FLAG\_SP\_LGA} &  60.4 & 73.8 \\
         \texttt{FLAG\_SP\_GC\_CONS} & 54.8 & 74.2 \\
         \texttt{FLAG\_NO\_RADEC\_ERR} & 40.7 & 81.0 \\
         \texttt{FLAG\_NO\_CTS\_ERR} & 37.9 & 81.1 \\
         \texttt{FLAG\_NO\_EXT\_ERR} & -- &    81.4 \\
         \texttt{FLAG\_OPT} & 73.6 & -- \\
    \hline
    \end{tabular}
\end{table}

We can estimate the increase in the spurious fraction in overdense regions and in groups of sources with other warning flags, using associations with the 5XMM-DR15 catalogue. For each warning flag, the percentages of eRASS:3 (main) detections having a strong association in 5XMM-DR15 are listed in Table \ref{tab:spurious_source_flag_assoc}. Only eRASS:3 detections with detection likelihoods \texttt{DET\_LIKE\_0}~$>10$ were used for this comparison. Lower association fractions indicate a higher rate of spurious detections in the flagged sub-group. We conclude that most flags indeed indicate an increased rate of spurious detections. However, the sources with flag \texttt{FLAG\_SP\_SCL} (stellar clusters) are mostly unproblematic. Also, extended sources with flags indicating missing error estimates do not suffer from increased contamination. Hence, a suitable strategy for selecting a sample of point sources with minimal contamination is to exclude all flagged sources except those with the \texttt{FLAG\_SP\_SCL} flag. The situation is more complex for extended sources. A major source of spurious extended detections is the fragmentation of large extended sources such as nearby galaxy clusters or supernova remnants. Most of these problematic sky areas are covered by the flagging scheme. However, the primary source of the large-scale extended emission also carries the respective flag. 

In Sect.~\ref{sec:erass3_catalogs}, we point out that the selection of sources only detected in the hard energy band ($2.3-5.0$~keV) leads to a significant spurious fraction. This fraction can be reduced by further filtering on the hard band detection likelihood. With a threshold of \texttt{DET\_LIKE\_3}~$=20$, close to $50\%$ of the selected hard-only sources have a counterpart in the DR1 hard catalogue. Since the eRASS1 catalogue is shallower than eRASS:3, this estimate is rather conservative.

In Sect.~\ref{sec:known_issues}, we note that in some cases, due to a confusion issue, relatively bright sources are not included in the catalogue. For previously unknown sources, there is obviously no workaround for this issue. However, for specific objects, catalogue users can search the DR1 catalogues for a detection and visually check the public DR1 image data to confirm the presence of a source.

%
%

%
\section{Multi-wavelength counterparts to eROSITA/DR2}
\label{sec:CTP_class}
Alongside the X-ray catalogues, we have also released a series of catalogues of optical/infrared counterparts (Sect.~\ref{subsec:ctp}) and classifications into Galactic or extragalactic sources (Sect.~\ref{subsec:class}), for the main and hard samples. For extragalactic sources (mostly AGNs), additional information, such as redshifts, will be presented in Roster et al. (in prep.), either from publicly available spectroscopic data or from dedicated photometric redshift computations. In all steps, we used the procedure described in \cite{Salvato2025}, and extensive details are provided there. The procedure is repeated for the eRASS:3 main and hard samples; below, we briefly summarise the steps and provide a general overview of the sources' properties in the eRASS:3 main catalogue. 

\begin{figure}
    \centering
    \includegraphics[width=0.45\textwidth]{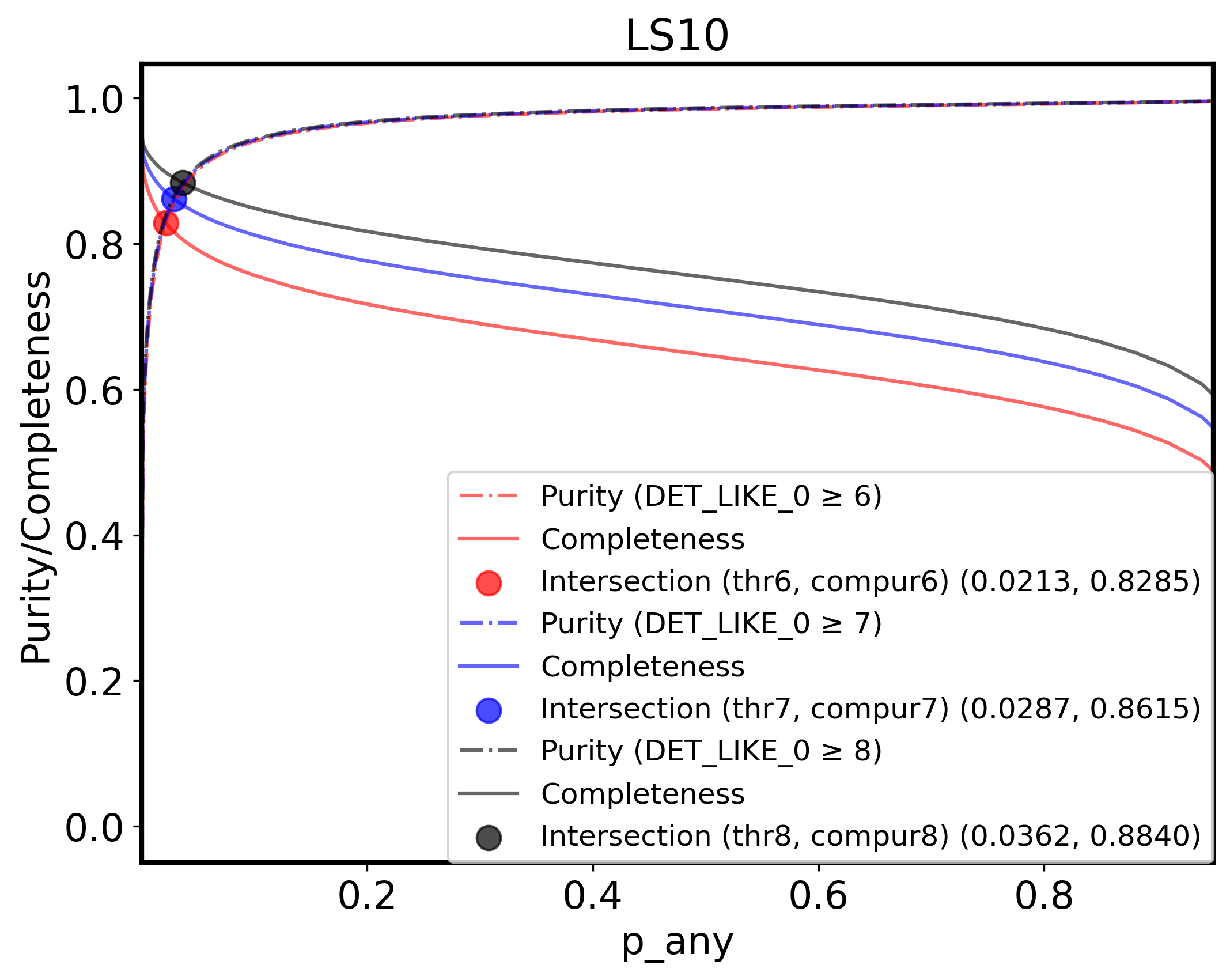}
    \includegraphics[width=0.45\textwidth]{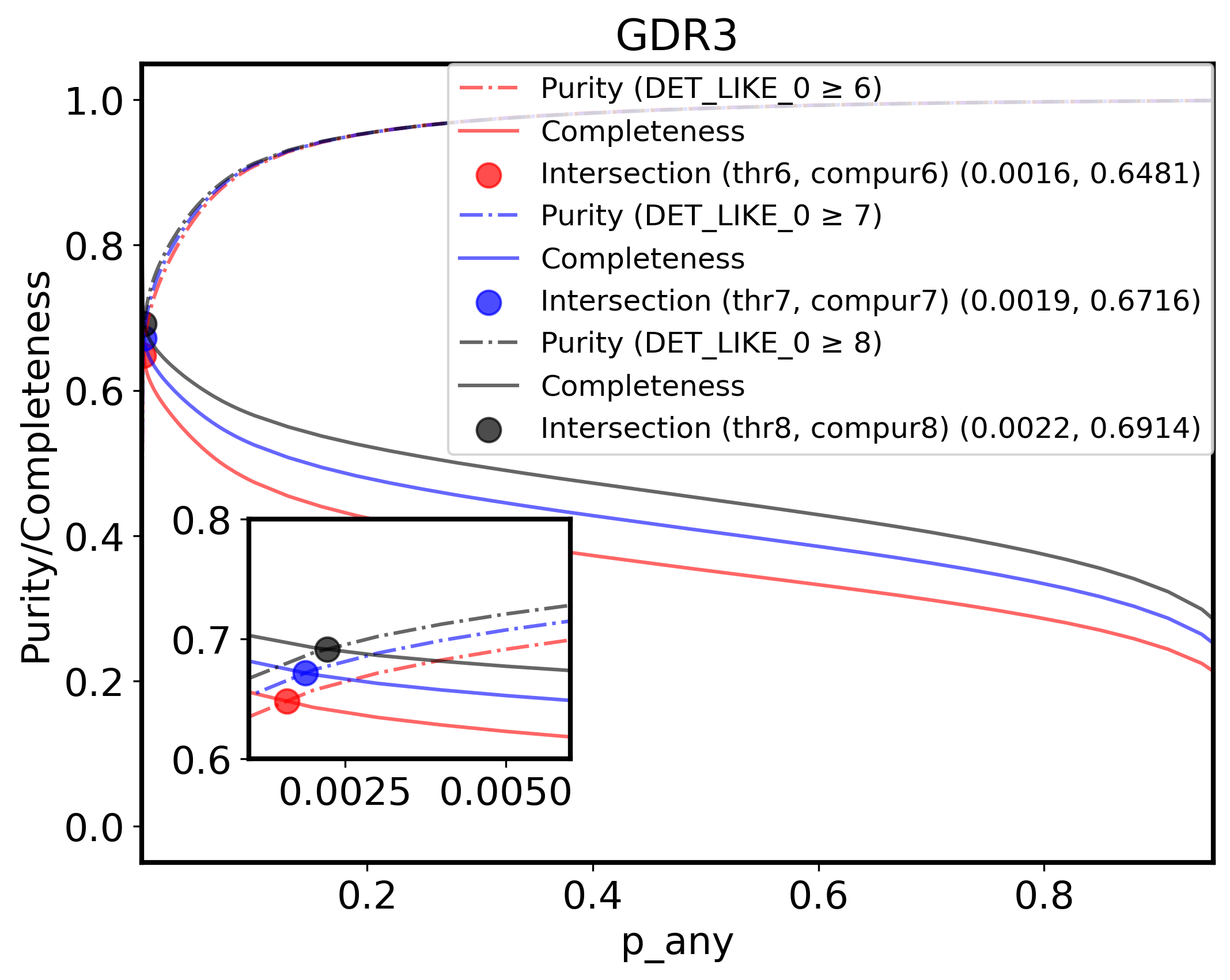}
    \includegraphics[width=0.45\textwidth]{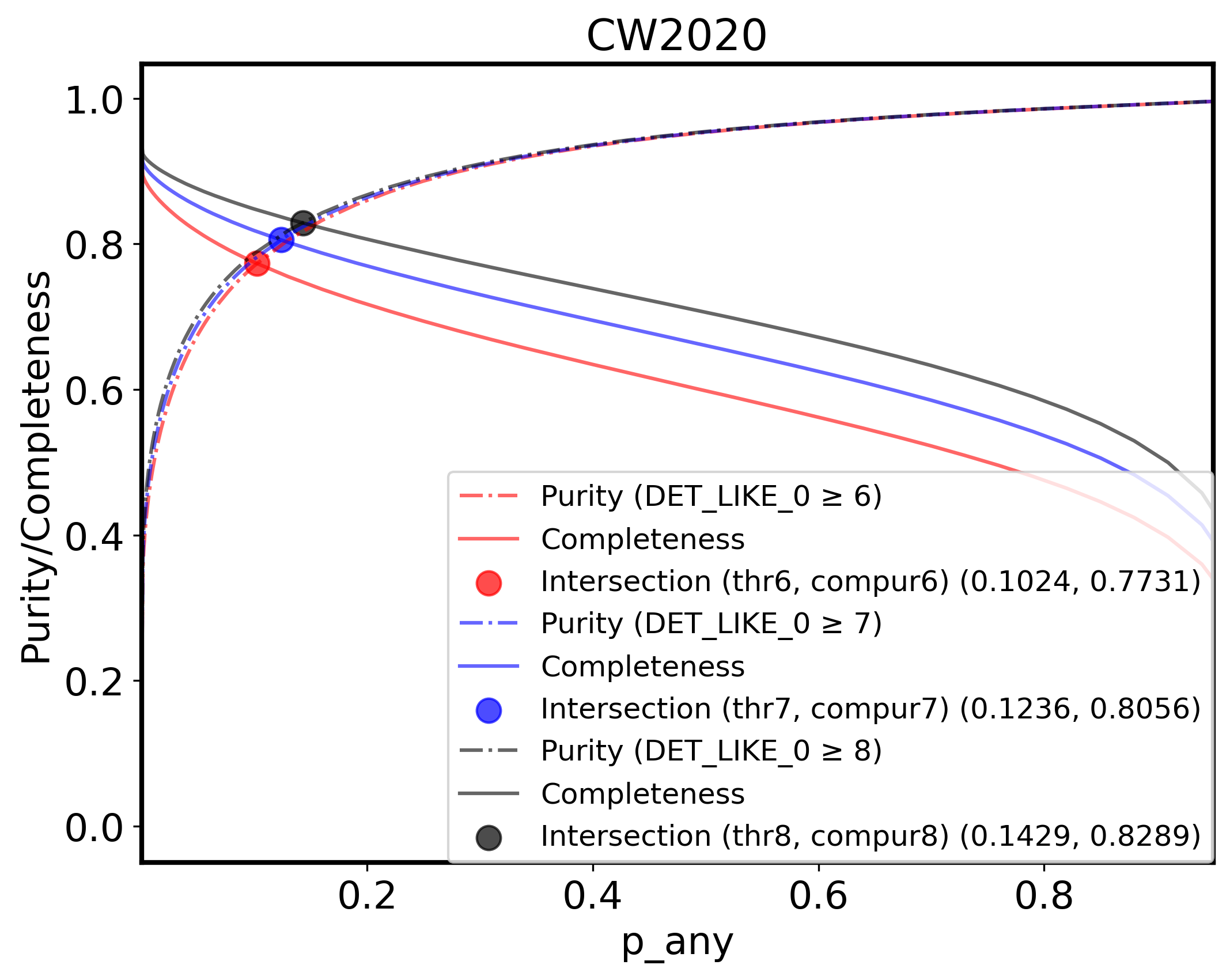}
    \caption{Cumulative purity (dashed lines) and completeness (solid lines) as a function of the \texttt{p\_any} threshold, computed for the entire sky using the ancillary data from LS10, GDR3, and CW2020, respectively. We also compute the values for three different \texttt{DET\_LIKE\_0}.}
    \label{fig:compur}
\end{figure}

\subsection{Counterpart identifications}
\label{subsec:ctp}
Multi-wavelength counterparts were identified by matching the X-ray catalogue to the LS10 \citep{Dey2019,Zenteno2025}, GDR3 \citep{gdr3}, and CatWise2020 \citep[CW2020;] [] {Marocco2021}, separately. As described in Sect.~3.1 of \cite{Salvato2025}, for cross-matching, we used the Bayesian algorithm NWAY\footnote{\url{https://github.com/JohannesBuchner/nway/}} \citep{Salvato2018}. We computed, for each optical/mid-infrared source (OMIR for short) within $60\arcsec$ from an X-ray position, the probability to be the right counterpart (\texttt{p\_i}), and the probability that the X-ray source has an OMIR counterpart (\texttt{p\_any}). These probabilities are computed considering $a)$ the angular separation from the X-ray source, $b)$ the positional errors, $c)$ the number density of the sources in the X-ray and OMIR catalogues (X-ray/OMIR), and $d)$ the a priori predicted probability of being an X-ray emitter from the OMIR catalogue information alone. This latter probability multiplication factor \citep[\texttt{bias\_Xproba}, see Sect.~3.1.1 in ][]{Salvato2025} is computed by comparing the spectral energy distributions (SEDs) of secure counterparts to 4XMM and {\it Chandra} sources with sources in the field that are not X-ray emitting (at least not at the depth of the observation),  using a random forest-based algorithm. This procedure is repeated for each auxiliary catalogue. The list of features adopted for each auxiliary catalogue, together with the prior's recall and leak fraction, is presented in Table 2 and Fig.~3 of \citet{Salvato2025}, respectively. For each eROSITA source, we identify the best counterpart with \texttt{match\_flag\,=\,1}, while we indicate with \texttt{match\_flag\,=\,2} the second best counterparts, if existing\footnote{The second best counterpart has a \texttt{p\_i} which differs from the \texttt{p\_i} of the best counterpart less than 0.05.}.

To quantify the probability that an X-ray/OMIR association is due solely to chance (random associations), we shifted the coordinates of the eROSITA sources by $3\arcmin$ in RA and Dec and repeated the association analysis with the same prior. In the vast majority of cases, the counterpart association probabilities \texttt{p\_any} are significantly lower than with the actual X-ray coordinates \citep[see Fig.~5 in][]{Salvato2025}. Finally, to estimate the completeness or purity at a given \texttt{p\_any} value, we compute the number of sources with higher \texttt{p\_any} in the actual catalogue (giving completeness) or shifted catalogue (giving spurious associations) \citep[see Fig.~6 of][]{Salvato2025}.

The results for the entire sky are illustrated in Fig.~\ref{fig:compur}, for LS10 (top panel), GDR3 (middle panel), and CW2020 (bottom panel), respectively. The purity and completeness (and thus the reliability) are higher for LS10 than for CW2020 and GDR3, because of the depth of the optical data and the rich number of features used as priors, which drastically reduces the effective number of sources that could become counterparts \citep[see also][]{Salvato2025}. 
Choosing different X-ray detection reliability thresholds on \texttt{DET\_LIKE\_0} slightly changes the results. Three different thresholds for \texttt{DET\_LIKE\_0\,>\,6,7,8} (red, blue, black) are presented in each panel of Fig.~\ref{fig:compur}. 
The completeness curves increase with higher X-ray detection reliability. This is because when the fraction of spurious X-ray detections increases, the \texttt{p\_any} probabilities become more similar to probabilities computed at random coordinates. Therefore, at higher \texttt{DET\_LIKE\_0} cuts, higher counterpart purity can be achieved.

The dots in Fig.~\ref{fig:compur} identify the \texttt{p\_any} value where purity (dot-dashed curve) and completeness (solid curve) are equal and balanced. They balance at slightly lower values than for eRASS1 \citep[see Fig.~6 of][]{Salvato2025}, but this is expected: given the increased depth of the eRASS:3 X-ray catalogue, the supporting multi-wavelength data might not always be deep enough to identify the correct counterpart. This is particularly true for GDR3 (middle panel of Fig.~\ref{fig:compur}), where the balancing is happening at an extremely low \texttt{p\_any}, and the purity and completeness are significantly lower than for CW2020 and LS10. This indicates that the GDR3 catalogue is too shallow for the depth of eRASS:3. Looking forward to the final full-depth (eRASS:5) eROSITA data release, expected in 2028, we will update the counterpart training data with LSST \citep{Ivezic2019}.

Given that the OMIR catalogues have different depths across the sky, and the OMIR source density also varies across the sky, a high purity, high completeness, and purity-completeness parity threshold on \texttt{p\_any} is calibrated in each sky tile, with the procedure described above, as described also in \citet{Salvato2025}. For each source, and given a \texttt{DET\_LIKE\_0} cut (6,7,8), we provide in the counterparts catalogues as \texttt{threshold[6,7,8]}, and the \texttt{p\_any} cut that balances purity and completeness, so that \texttt{threshold[6,7,8]>p\_any} gives good results over the entire sky. 
Conversely, for each source, we provide the purity and completeness corresponding to the specific \texttt{p\_any} in that tile, at the \texttt{DET\_LIKE\_0} of the source.
These values are indicated as \texttt{comp[6,7,8]} and \texttt{pur[6,7,8]}.

Summarising, for each X-ray source, and each auxiliary catalogue (\texttt{aux}\,=\,LS10, GDR3, or CW2020, with corresponding column suffix \texttt{\_aux}), we provide:
\begin{itemize}
\item \texttt{NWAY\_p\_any}: The probability that the X-ray source has a counterpart in the auxiliary catalogue.
\item \texttt{NWAY\_p\_i}: For each OMIR source within 1 arcmin from the X-ray position, the relative probability of the combination of OMIR - X-ray source match \citep[see][]{Salvato2018}.
\item \texttt{NWAY\_bias\_Xproba}: The probability multiplication factor from OMIR SED information.
\item \texttt{NWAY\_completeness[6,7,8]}: The completeness corresponding to the particular \texttt{NWAY\_p\_any} in that tile.
\item \texttt{NWAY\_pur[6,7,8]}: The purity corresponding to the particular \texttt{NWAY\_p\_any} in that tile.
\item \texttt{NWAY\_threshold[6,7,8]}: The \texttt{p\_any} threshold where purity and completeness are equal, for each parent sample cut \texttt{DET\_LIKE\_0$>[6,7,8]$}, specific to the tile of this object.
\item \texttt{NWAY\_compur[6,7,8]}: the value of completeness and purity at the \texttt{NWAY\_threshold[6,7,8]}.
\end{itemize}

In this way, the user can fully control the selection function when they construct their samples of interest. An obvious question at this point is whether there is consistency among the counterparts selected using LS10, GDR3, and Cw2020. Table \ref{tab:GDR3_CW2020_inAnyLS10} shows that in the footprints of LS10, where all the optical bands reach the nominal depth \citep[\texttt{inAllLS10}; see][]{Salvato2025}, the agreement between LS10 and GDR3 is very low (around 50\%, depending on the {\texttt{DET\_LIKE\_0} value). The fact that limiting the LS10 counterparts to the same magnitude limit as GDR3 increases the agreement to $92-93\%$ indicates that, overall, about 50\% of the GDR3 counterparts are chance associations, and the correct counterparts are actually fainter and undetected by GDR3. To overcome this problem, we suggest that the user rely on the GDR3 counterparts only if their \texttt{p\_any} has high purity. For example, 50\% of the eRASS:3 sources in GDR3 have \texttt{p\_any}, which corresponds to \texttt{NWAY\_pur[6]}\,>\,90\%, and 90\%  of those have the same counterpart in GDR3 and LS10. The comparison between LS10 and CW2020 is much better, ranging from 81.4\% to 86.8\%, and improves further when we consider the LS10 counterparts at CW2020's magnitude limit.
In summary, the user should use the LS10 counterparts when possible and rely on GDR3 and Cw2020 counterparts only if \texttt{p\_any} corresponds to high purity (e.g. \texttt{NWAY\_pur[6,7,8]}\,>\,0.9).

\begin{table}
\centering 
\small
\caption{Fraction of sources in GDR3 and CW2020 sharing the same counterpart as identified using LS10 for the \texttt{inAllLS10} region.} 
\renewcommand{\arraystretch}{1.2}
\begin{tabular}{c|r|c|c}
\hline
\hline
\multirow{2}{*}{\texttt{DET\_LIKE\_0}} & 
\multirow{2}{*}{\texttt{inAllLS10}} & 
\multicolumn{2}{c}{same CTP} \\
&  &  LS10-GDR3  &  LS10-CW2020 \\
   \hline
   $\ge6$ & \num{1280090} & 48.7\,\% & 81.4\,\%  \\
   $\ge6$ \& {\it g}\,<\,20.7 & \num{548050} & 91.8\,\% & -- \\
   $\ge6$ \& $W1<20.4$ & \num{1069737} & -- & 89.2\,\%\\
   \hline
   $\ge7$ & \num{1067816} & 53.3\,\%& 84.5\,\% \\
   $\ge7$ \& $g<20.7$ & \num{498814} &92.7\,\% & -- \\
   $\ge7$ \& $W1<20.4$ & \num{920525} &--&90.1\,\% \\
   \hline
   $\ge8$ & \num{920846} & 56.9\,\% &86.8\,\%  \\
   $\ge8$ \& $g<20.7$ & \num{460168} & 93.3\,\%& -- \\
   $\ge8$ \& $W1<20.4$ & \num{811210} & --  & 92.0\,\%\\
\hline
\end{tabular}
\tablefoot{Only the first best counterpart (\texttt{match\_flag}\,=\,1) is considered in this comparison, and no cut on \texttt{p\_any} is applied.}
\label{tab:GDR3_CW2020_inAnyLS10}
\end{table}

\subsection{Galactic and extragalactic classification}
\label{subsec:class}
In this section, we describe how we classify the matched counterparts to the eRASS:3 sources as Galactic or extragalactic. This follows the procedure outlined in Sect.~5 of \cite{Salvato2025}, which depends on the auxiliary catalogues. For the counterparts determined in LS10, we combine four approaches. First, we use STAREX, a machine learning algorithm developed by our group \citep[see details in Sect.~5.3.1 of][]{Salvato2025}, to determine the probability of being Galactic or extragalactic. We used Eqs.~2 and 3 from \citet{Salvato2025}, depending on whether Gaia astrometric information was available for the source. Then we considered the Galactic-extragalactic line separators presented in \citet{Salvato2018} (using X-ray flux and W1 magnitude)  and \citet{Salvato2022} (using {\it g$-$r} and {\it z$-$W1} colours). Each method is reliable depending on data availability, and thus, we provide the classification from each method. Additionally, we provide the \texttt{class\_gal\_exgal} column \citep[described in more detail in Sect.~5.3.2 in][]{Salvato2025}, which consolidates the results in the following order:
\begin{itemize}
\item \texttt{class\_gal\_exgal}\,$=5$: prob\_STAREX>95\% (highly secure extragalactic).
\item \texttt{class\_gal\_exgal}\,$=4$: prob\_STAREX>50\% (secure extragalactic).
\item \texttt{class\_gal\_exgal}\,$=3$: prob\_STAREX<50\% and classified as extragalactic both by the \citet{Salvato2018} \& \citet{Salvato2022} methods;
\item \texttt{class\_gal\_exgal}\,$=2$: prob\_STAREX<50\% and classified as extragalactic by either \citet{Salvato2018} or \citet{Salvato2022} methods;
\item \texttt{class\_gal\_exgal}\,$=1$: extended in LS10.
\item \texttt{class\_gal\_exgal}\,$=-1$: to the remaining sources, assumed to be stars.
\item \texttt{class\_gal\_exgal}\,$= -5$: secure stars. These are point-like in LS10 and with highly significant ($5\sigma$) parallax or proper motion\footnote{\texttt{SNR\_parallax}\,$>5$ or \texttt{SNR\_PM}\,$>5$; see Sect.~5.3.2 of \citet{Salvato2025}}.
\end{itemize}
Additionally, we identify known blazars, flat-spectrum radio sources, X-ray binaries, cataclysmic variables, and high-mass or low-mass X-ray binaries from archival data (i.e. Simbad) cross-matched with the optical counterpart within $1\arcsec$. The information is given in the columns \texttt{class\_jetted} and \texttt{simbad\_known\_galactic} \citep[see also last paragraph of Sect.~5.3.2 in][]{Salvato2025}. Following this scheme, \num{1195316} (88\%) of the LS10 counterparts are considered extragalactic. Among these, 84\% have less than 10\% probability to be chance associations (NWAY\_purity6>0.9).

\begin{table*}
\centering
\caption{Number of Galactic (mostly stars) and extragalactic (mostly AGN) sources in the overlap region between eRASS:3 and LS10, focusing on purity, completeness, and their trade-off. See end of Section~\ref{subsec:examples} for additional details.}
\renewcommand{\arraystretch}{1.2}
\label{tab:LS10classes}
\begin{tabular}{p{8cm} | c c c}
\hline
\hline
Parent sample & & Number & \\
\hline
X-ray point sources, no flags, including multiple counterparts, within LS10 area$^1$
(sources with multiple counterparts are repeated) & \multicolumn{3}{c}{\num{1475473}} \\
 & \\
With at least a counterpart$^2$ & \multicolumn{3}{c}{\num{1373804}} \\
\\
 & \multicolumn{3}{c}{Example counterpart selection strategy} \\
Counterpart sample & Complete & Pure & Balanced \\
\hline
\hline
Best counterpart & \,\num{1095679}$^3$ (100\%) & \,\num{1042307}$^4$ (100\%) & \,\,\,\num{1001917}$^5$ (100\%) \\
Extragalactic (AGNs)$^6$  & \num{966996} (88\%)  & \num{930532} (89\%) & \num{902743} (89\%) \\
Galactic (stars)$^7$ & \num{128683} (12\%) & \num{111775} (11\%) & \num{99174} (11\%) \\
\hline
\end{tabular}
\tablefoot{
Applied selections:\\
$^1$: \texttt{EXT=0 \& FLAG\_SP\_\{SNR,BPS,SCL,LGA,GC\_CONS\}==0 \& FLAG\_NO\_\{RADEC,EXT,CTS\}\_ERR=0}  \\
$^2$: \texttt{NWAY\_ncat=2 \& NWAY\_match\_flag=1}. \\
$^3$: \texttt{NWAY\_match\_flag=1 \& NWAY\_completeness6<0.8}. \\
$^4$: \texttt{NWAY\_match\_flag=1 \& NWAY\_purity6>0.9 \& NWAY\_p\_i>0.9} (the last removes ambiguous situations). \\
$^5$: \texttt{NWAY\_match\_flag=1 \& NWAY\_p\_any>NWAY\_threshold6 \& NWAY\_compur6>0.85} (note: this can result in incomplete sky coverage). \\
$^6$: additionally \texttt{class\_gal\_exgal>0}. \\
$^7$: additionally \texttt{class\_gal\_exgal<0}.
}
\end{table*}

For classifying CW2020 and GDR3 sources, the approach defined in Sects.~5.2. and 5.3. of \citet{Salvato2022} is followed. Briefly, for GDR3, we do compute classification, but provide the information from the Gaia collaboration, summarised in the quantities \texttt{PQSO}, \texttt{PGAL}, \texttt{Pstar}, \texttt{PWD}, and \texttt{Pbin}, which represent the probability of being a QSO, a galaxy, a star, a white dwarf, or a binary, respectively \citep{gdr3}. In GDR3, we define extragalactic sources as those with \texttt{PGAL}~$>0.8$ or \texttt{PQSO}~$>0.8$. Based on this definition, among the \num{951335} sources that have \texttt{NWAY\_purity6}>0.9 (see Section \ref{subsec:ctp} for the reason of this selection), \num{392166} (41.2\%) are considered of secure extragalactic nature.

For CW2020, {\it W1$-$W2} $>0.3$ in the Vega system identifies extragalactic sources, with minimal contamination from Galactic sources. However, a non-negligible fraction of sources with {\it W1$-$W2} $<0.3$ are extragalactic \citep[see Fig.~13 of][]{Salvato2025}, and this fraction varies with latitude and depth. This cut yields a pure but incomplete sample of \num{1269265} extragalactic sources out of \num{1832856} (69.3\%). Further refinements are left to the user.

\subsection{User case examples and best practice}
\label{subsec:examples}
In the section above, we have described how the catalogues of counterparts to eROSITA-DE DR2 are made. We provide counterparts using LS10, CW2020 and GDR3, for the main and hard point sources, for a total of 6 catalogues. The catalogues can serve many purposes; below, we provide use-case examples and best-practice suggestions.
\begin{itemize}
    \item The user wants to check whether some sources of interest are associated with an X-ray detection. For that, the user has to be careful to match the coordinates using the counterpart's coordinates rather than the X-ray source's, since, given the resolution, the associated counterpart can be many arcseconds away. The coordinates of the counterparts are given as [LS10, CW2020, GDR3]\_[RA, DEC]. After identifying the corresponding eROSITA source, the user must perform an internal match in the catalogue using the X-ray source ID to ensure the counterpart is unique. This is because, in some cases, there is more than a possible counterpart, suggesting that the X-ray flux might be the sum of the contributions from many sources. In cases where counterparts are identified in the GDR13 and CW2020 catalogues, the user should also examine the value assigned to \texttt{NWAY\_purity[6,7,8]}, depending on the \texttt{DET\_LIKE\_0} of the source. This value tell what is the probability of a chance association.  For example, for Gaia counterparts, most associations are due to chance, unless \texttt{NWAY\_purity[6,7,8]}~$>0.9$. (see previous section).\\
    \item The user is interested in creating a sample of eROSITA Galactic/extragalactic sources with a well-defined selection function. Given the different depth and source density of the eROSITA and OMIR data across the sky, the same value of \texttt{p\_any} or \texttt{p\_i} (see definition in Sect.~\ref{subsec:ctp}) has a different value of purity and completeness. Instead, to create samples that control for purity and/or completeness, the user should define the value \texttt{NWAY\_purity[6,7,8]} or \texttt{NWAY\_completeness[6,7,8]}. Alternatively, the user can decide to maximise completeness and purity above a certain level across the entire sky. In that case, the column to look for is \texttt{NWAY\_compur[6,7,8]}. The user can then select extragalactic (or Galactic) sources using \texttt{class\_gal\_exgal}>(<)0.
    Table~\ref{tab:LS10classes} illustrates an example for LS10 when pursuing (1) an inclusive selection, optimised for 80\% counterpart completeness; (2) a pure selection, optimised for 90\% counterpart purity and restricted to unambiguous counterparts (\texttt{p\_i > 0.9}); and (3) a balanced selection, optimised to maximise both purity and completeness for the tile while maintaining values of at least 85\% for each.
\end{itemize}
These are just examples; further refinements are possible. A Jupyter Notebook released with eROSITA-DE DR1 counterparts via Zenodo\footnote{\url{https://zenodo.org/records/17404798}} \citep[see details in][]{Salvato2025}, can be easily adapted for DR2, given the similarity in the names of the columns.

%
%
%

%
\section{eROSITA-DE Data Release 2 (DR2)}
\label{sec:dr2_description}
The eROSITA-DE Data Release 2\footnote{\url{https://erosita.mpe.mpg.de/dr2}} (DR2) primarily consists of the main and hard source catalogues described in Sect.~\ref{sec:erass3_catalogs}. In contrast to the Early Data Release (EDR) and DR1, DR2 does not include data (i.e. photon event lists, source products, etc.) from the Calibration and Performance Verification (CalPV) phase nor the eRASS:3 survey period. Similarly, the version of the eSASS software used to process and analyse eRASS:3 data is not included in this release. 

DR2 provides the eROSITA upper-flux-limit server \citep{Tubin2024}, as in DR1. Upper limits are derived from standard eROSITA calibration products (counts, background, and exposure maps) using the Bayesian method of \cite{Kraft1991}. Pre-computed $3\sigma$ ($99.87\%$) upper limits are available at every sky pixel in the eROSITA-DE footprint, stored in HEALPix format for fast querying. Products are delivered for the 1B band ($0.2-2.3$~keV) and for the hard 3B energy band ($2.3-5.0$~keV). The data can be accessed either by downloading the full pre-computed set or via an online web tool\footnote{\url{https://erosita.mpe.mpg.de/dr2/AllSkySurveyData_dr2/UpperLimitServer\_dr2}}.

Together with the eRASS:3 main and hard catalogues, DR2 also includes six counterpart catalogues, computed for the main and hard samples, using LS10, GDR3, and CW2020, respectively (see Sect.~\ref{sec:CTP_class}). The catalogues include probabilities of association, classifications, and basic information on the counterparts (e.g. magnitudes). The ID from the original ancillary catalogue is also provided, making it trivial to retrieve additional properties. The six catalogues only include those DR2 sources with a counterpart and thus they are subsamples of the DR2 catalogues.

%
%

%
\section{Summary and outlook}
\label{sec:summary_outlook}
This work presents the operations, observations, data reduction, and analysis of the first 1.5 years of the eROSITA all-sky survey (eRASS). It also introduces the resulting X-ray source catalogues, which are based on three complete all-sky scans (eRASS:3) and cover the western Galactic hemisphere under the proprietary rights of the eROSITA German Consortium. We present the catalogues of X-ray sources extracted using the standard eSASS pipeline and describe the quality-control tests applied to the catalogued sources, including assessments of astrometric accuracy, photometric reliability, and catalogue fidelity. We also present the catalogues produced using a systematic approach to link eROSITA point sources to their most reliable counterparts in optical (LS10 and GDR3) and mid-infrared (CW2020) data and to determine their Galactic/extragalactic nature. Finally, we also describe the content of the data release associated with eRASS:3 to facilitate the scientific exploitation of these rich catalogues by the broader astronomical community.

The scientific and informational content of the eRASS:3 survey is extensive and diverse, making it difficult to condense into a brief summary. Therefore, the focus of this work lies primarily on the eRASS:3 X-ray catalogues, the properties of the sources contained therein, along with their counterpart identifications. Among the various investigations made possible by the eROSITA-DE DR2, we have made particular note of the description of the SDSS-V spectroscopic follow-up program (BHM/SPIDERS, Merloni et al., in prep.), the study of the AGN X-ray luminosity function evolution (Roster et al., in prep., which will also release photometric and spectroscopic redshift compilations for the eRASS:3 AGNs), the systematic analysis of AGN high-amplitude and Blazar variability (Bahic et al., in prep.; H\"ammerich et al., in prep.), and the characterisation of the diverse X-ray-emitting stellar populations of the Milky Way, from CVs (Brink et al., Schwope et al., in prep.) to main sequence stars (Robrade et al., in prep.) to dwarfs (Stelzer et al., in prep.).

The eRASS:3 X-ray catalogues mark a substantial expansion over eRASS1, featuring more than twice as many point-like and extended sources. Because of this growth, most detections represent previously unknown X-ray emitters, underscoring the transformative impact of the eRASS:3 half-sky catalogues on our view of the X-ray Universe. Their long-term scientific value is further amplified by the wealth of complementary all-sky and wide-area surveys available at other wavelengths. In particular, optical and infrared programmes such as Gaia, SDSS, Pan-STARRS, DES, the Legacy Survey, HSC, VISTA/VHS, and WISE have already demonstrated how multi-wavelength analyses can significantly deepen our understanding of high-energy astrophysical phenomena. 

The eROSITA-DE DR3 is planned for 2028 and will make publicly available all eROSITA data acquired during eRASS:5 in the western Galactic hemisphere. In addition to the source catalogues, the release will include associated analysis software and higher level data products. It is expected that the future released data will be processed with a new version of the eSASS software, including an updated calibration database.
This final eROSITA data release will build on this foundation by combining observations from multiple all-sky passes. Beyond delivering increasingly deep and uniform coverage of the X-ray sky, these releases will open the door to systematic investigations of source variability on timescales from months to years. With the current level of data quality and continued calibration refinements, the next release is expected to yield high-fidelity catalogues comprising several million X-ray sources, in line with the performance forecasts established in the early stages of the eROSITA mission.

\section{Data availability}
The catalogues presented in this paper are publicly available through the eROSITA DR2 web pages at \url{https://erosita.mpe.mpg.de/dr2/AllSkySurveyData_dr2/Catalogues_dr2/}. They are also available from the CDS via \url{cdsarc.cds.unistra.fr}.

\bibliographystyle{aa}
\bibliography{references}

%
%

\begin{appendix} 
\section{Acknowledgements}
\label{appendix:acknowledgement}
\begin{acknowledgement}
The authors thank the referee for their helpful and constructive comments on the draft. \\

This work is based on data from eROSITA, the soft X-ray instrument aboard SRG, a joint Russian-German science mission supported by the Russian Space Agency (Roskosmos), in the interests of the Russian Academy of Sciences represented by its Space Research Institute (IKI), and the Deutsches Zentrum f{\"{u}}r Luft und Raumfahrt (DLR). The SRG spacecraft was built by the Lavochkin Association (NPOL) and its subcontractors and is operated by NPOL with support from the Max Planck Institute for Extraterrestrial Physics (MPE).\\

The development and construction of the eROSITA X-ray instrument were led by MPE, with contributions from the Dr. Karl Remeis Observatory Bamberg \& ECAP (FAU Erlangen-Nuernberg), the University of Hamburg Observatory, the Leibniz Institute for Astrophysics Potsdam (AIP), and the Institute for Astronomy and Astrophysics of the University of T{\"{u}}bingen, with the support of DLR and the Max Planck Society. The Argelander Institute for Astronomy of the University of Bonn and the Ludwig Maximilians Universit{\"{a}}t Munich also participated in the science preparation for eROSITA.\\

The eROSITA data shown here were processed using the eSASS/NRTA software system developed by the German eROSITA consortium. \\

W. Becker acknowledges support from the Deutsche Forschungsgemeinschaft through the project eROSTEP (BE 1649/11-2) within the Research Unit FOR 2990 (eRO-STEP).\\

E. Bulbul, E. Artis, S. Zelmer, and X. Zhang acknowledge financial support from the European Research Council (ERC) Consolidator Grant under the European Union’s Horizon 2020 research and innovation program (grant agreement CoG DarkQuest No 101002585).\\

J.V. Corr\^{e}a-Rodrigues acknowledges support from the S\~{a}o Paulo Research Foundation (FAPESP), Brazil. Process Numbers \#2022/09374-0, and \#2025/03535-0.\\

G. P\"uhlhofer acknowledges support from the Deutsche Forschungsgemeinschaft through the project PU 308/2-2 within the Research Unit FOR 2990 (eRO-STEP).\\

M. Sasaki acknowledges support from the Deutsche Forschungsgemeinschaft through the projects SA 2131/13-2, SA 2131/14-2, and SA 2131/15-2 within the Research Unit FOR 2990 (eRO-STEP).\\

S. Saedi acknowledges support from the Deutsche Forschungsgemeinschaft through the project  DFG SA 4388/2-1 within the Research Unit FOR 2990 (eRO-STEP).\\

V. Suleimanov acknowledges support from the Deutsche Forschungsgemeinschaft  (grant WE 1312/59-1).\\

This research has made use of data obtained from the Chandra Source Catalogue, provided by the Chandra X-ray Center (CXC).\\

This research has made use of the SIMBAD database \cite{Wenger2000} and the VizieR catalogue access tool \cite{Ochsenbein2000vizier}, operated at CDS, Strasbourg, France.\\

This research uses services or data provided by the Astro Data Lab at NSF's National Optical-Infrared Astronomy Research Laboratory. NOIRLab is operated by the Association of Universities for Research in Astronomy (AURA), Inc., under a cooperative agreement with the National Science Foundation. \\

The Legacy Surveys consist of three individual and complementary projects: the Dark Energy Camera Legacy Survey (DECaLS; Proposal ID \#2014B-0404; PIs: David Schlegel and Arjun Dey), the Beijing-Arizona Sky Survey (BASS; NOAO Prop. ID \#2015A-0801; PIs: Zhou Xu and Xiaohui Fan), and the Mayall z-band Legacy Survey (MzLS; Prop. ID \#2016A-0453; PI: Arjun Dey). DECaLS, BASS and MzLS together include data obtained, respectively, at the Blanco telescope, Cerro Tololo Inter-American Observatory, NSF’s NOIRLab; the Bok telescope, Steward Observatory, University of Arizona; and the Mayall telescope, Kitt Peak National Observatory, NOIRLab. Pipeline processing and analyses of the data were supported by NOIRLab and the Lawrence Berkeley National Laboratory (LBNL). The Legacy Surveys project is honoured to be permitted to conduct astronomical research on Iolkam Du’ag (Kitt Peak), a mountain with particular significance to the Tohono O’odham Nation.\\

NOIRLab is operated by the Association of Universities for Research in Astronomy (AURA) under a cooperative agreement with the National Science Foundation. LBNL is managed by the Regents of the University of California under contract to the U.S. Department of Energy.\\

This project used data obtained with the Dark Energy Camera (DECam), which was constructed by the collaboration of the Dark Energy Survey (DES). Funding for the DES Projects has been provided by the U.S. Department of Energy, the U.S. National Science Foundation, the Ministry of Science and Education of Spain, the Science and Technology Facilities Council of the United Kingdom, the Higher Education Funding Council for England, the National Center for Supercomputing Applications at the University of Illinois at Urbana-Champaign, the Kavli Institute of Cosmological Physics at the University of Chicago, Center for Cosmology and Astro-Particle Physics at the Ohio State University, the Mitchell Institute for Fundamental Physics and Astronomy at Texas A\&M University, Financiadora de Estudos e Projetos, Fundacao Carlos Chagas Filho de Amparo a Pesquisa do Estado do Rio de Janeiro, Conselho Nacional de Desenvolvimento Cientifico e Tecnologico and the Ministerio da Ciencia, Tecnologia e Inovacao, the Deutsche Forschungsgemeinschaft and the Collaborating Institutions in the Dark Energy Survey. The Collaborating Institutions are Argonne National Laboratory, the University of California at Santa Cruz, the University of Cambridge, Centro de Investigaciones Energeticas, Medioambientales y Tecnologicas-Madrid, the University of Chicago, University College London, the DES-Brazil Consortium, the University of Edinburgh, the Eidgenossische Technische Hochschule (ETH) Zurich, Fermi National Accelerator Laboratory, the University of Illinois at Urbana-Champaign, the Institut de Ciencies de l’Espai (IEEC/CSIC), the Institut de Fisica d’Altes Energies, Lawrence Berkeley National Laboratory, the Ludwig Maximilians Universitat Munchen and the associated Excellence Cluster Universe, the University of Michigan, NSF’s NOIRLab, the University of Nottingham, the Ohio State University, the University of Pennsylvania, the University of Portsmouth, SLAC National Accelerator Laboratory, Stanford University, the University of Sussex, and Texas A\&M University. \\

BASS is a key project of the Telescope Access Program (TAP), which has been funded by the National Astronomical Observatories of China, the Chinese Academy of Sciences (the Strategic Priority Research Program “The Emergence of Cosmological Structures” Grant \#XDB09000000), and the Special Fund for Astronomy from the Ministry of Finance. The BASS is also supported by the External Cooperation Program of the Chinese Academy of Sciences (Grant \#114A11KYSB20160057), and the Chinese National Natural Science Foundation (Grant \#12120101003, \#11433005). \\

The Legacy Survey team makes use of data products from the Near-Earth Object Wide-field Infrared Survey Explorer (NEOWISE), which is a project of the Jet Propulsion Laboratory/California Institute of Technology. NEOWISE is funded by the National Aeronautics and Space Administration. \\

The Legacy Surveys imaging of the DESI footprint is supported by the Director, Office of Science, Office of High Energy Physics of the U.S. Department of Energy under Contract No. DE-AC02-05CH1123, by the National Energy Research Scientific Computing Center, a DOE Office of Science User Facility under the same contract, and by the U.S. National Science Foundation, Division of Astronomical Sciences under Contract No. AST-0950945 to NOAO. \\

Software: astropy \citep{AstropyCollaboration2013,AstropyCollaboration2018}, topcat \citep{topcat}, stilts (\url{https://www.star.bris.ac.uk}), Sciserver \citep[implemented at MPE, based on][]{SciserverJHU2020},  matplotlib \citep{matplotlib}, scipy \citep{scipy}.

\end{acknowledgement}

\section{Software and calibration versions used in this work}
\label{appendix:esass}
The eROSITA data presented in this work were processed from September 2023 to June 2024. The software and calibration versions used (pipeline configuration 030) are different from those of the earlier eROSITA Early Data Release (EDR, pipeline configuration 001), the eROSITA Data Release 1 (DR1, pipeline configuration 010), and the one from later work based on proprietary eROSITA data (pipeline configuration 020). \cite{Merloni2024} highlights the differences between 001, 010 and 020. 

Table~\ref{tab:esass_versions} summarises the updates on eSASS task versions introduced in version 030, while Table~\ref{table:caldb} lists the corresponding calibration files.

\begin{table*}[htp]
  \centering 
  \small
  \caption{DR2 eSASS task versions and description of the main updates with respect to DR1 and 020 processing.}
  \renewcommand{\arraystretch}{1.2}
  \begin{tabular}{m{2.5cm}m{1.5cm}m{2cm}m{10.5cm}}
  \hline
  Pipeline module & eSASS task & Task version & Important functional improvements \\
  \hline
    & {\tt frameprep}  & 4.13.3 & Improves science frame quality checks and fits a time model to correct timestamp jitter. \\
TEL & {\tt evprep}     & 4.13.3 &  Now focused solely on event processing, the task reads {\tt frameprep} output, applies bug fixes, handles bad pixels, correctly adds leap seconds, omits MIP flagging, and includes a BADFRAMn extension in the output.\\
    & {\tt hotpixfind} & 1.11   & Routine limited to copying bad pixel entries from the calibration database. Its full functionality - to detect over-active pixels and update the BADPIX table—was not activated in 030 processing. \\
    & {\tt pattern}    & 4.3.9  & Adapted to new flagging scheme for bad pixels. \\
    & {\tt energy}     & 3.2.8  & Improved energy calibration; see Sect.~\ref{sec:energy_calibration}.  \\
    & {\tt evatt}      & 1.21.1 & Implemented time dependent astrometric correction. \\
  \hline
EXP & {\tt evtool}     & 2.39   & Now accepts either 'STDGTI' or multiple TM-specific 'STDGTIn' extensions. GTI filtering is more robust, now cutting holes for CORRUPT\_FRAME, OUT\_OF\_GTI, and ARTEFACT flags (using BADFRAMn if available), and supports filtering by owner and sky coordinates. \\ 
    & {\tt flaregti}   & 1.29   & No longer writes GTIs to event files; the output GTI file (gtifile parameter) is now mandatory and contains a single STDGTI extension with GTIs common to all TMs. \\
  \hline
DET & {\tt expmap}     & 2.24.9 &  The parameter interface is simplified—filtering now follows input event list keywords, with no user override. It supports owner and RA/DEC selection, prints expected exposure ratios, runs faster, and uses more robust attitude interpolation. \\
    & {\tt ermldet}    & 1.60.2 & Increased max iterations from 100 to 1000 and set a 0.05 minimum for positional parameters, improving fits for bright, slowly converging sources.  \\
 \hline   
SOU & {\tt srctool}    & 1.87   & The gtitype task option has been removed, and the GTI type is now read from the GTI\_SEL keyword. \\
\hline
  \end{tabular}
\label{tab:esass_versions}
\end{table*}

\begin{table*}[htp]
  \centering
  \small
  \caption{DR2 calibration files updated with respect to DR1 and 020 processing.}
  \renewcommand{\arraystretch}{1.2}
  \begin{tabular}{lcl}
  \hline
  Calibration component & Version DR2 & Comments \\
  \hline
  {\tt gyro\_boresight\_190714v<x>} & 12 & Boresight quaternion\\
  {\tt gyro\_boresight\_200611v<x>} & 11 & Boresight quaternion\\
  {\tt scc\_tdrift\_190701v<x>} & 2 & Reset times of and samples of fit to S/C clock\\
  {\tt srg\_tposcorr\_190611v<x>} & 1 & Time dependent positional correction\\
  {\tt tm<x>\_bad\_frames\_190701v<x>}  & 1/1/1/1/2/1/1 & Intercalated frames\\
  {\tt tm<x>\_badpix\_190712v<x>}  & 1/2/1/1/1/1/2 & eROSITA bad pixel list, supersedes all previous bad pixel lists\\
  {\tt tm<x>\_corr\_algos\_190701v<x>} & 1/-/-/-/2/-/2 & Algorithms for event corrections\\
  {\tt tm<x>\_detmap\_100602v<x>}  & 3 & eROSITA detector map\\
  {\tt tm<x>\_energy\_190714v<x>} & 7 & parameters for the CTI and Gain correction\\
  {\tt tm<x>\_mipneigh\_990101v<x>}$^a$ & 1 & Ratio of excluded MIP-neighbours to MIPs\\
  {\tt tm<x>\_pattern\_191023v<x>}  & 3 & split thresholds corresponding to 60 eV\\
  {\tt tm<x>\_resets\_190701v<x>}  & 1 & CE reset times\\
  {\tt tm4\_thrscal\_<date><version>} & 191001v01 & Scaled threshold map\\
  {\tt tm<x>\_timeoff\_190701v<x>} & 3/2/2/2/2/2/2 & Time offset between CE and ITC clocks\\
  \hline
  \end{tabular}
    \tablefoot{
    Different versions are separated by a slash. In case a calibration file does not exist for a TM it is marked by a dash.
    
    $^a$ {\tt tm<x>\_mipneigh\_990101v<x>} is missing in Table E.2 in \cite{Merloni2024} and is therefore listed here for completeness.
    
      \label{table:caldb}}
\end{table*}

\section{X-ray catalogue column descriptions}
\label{appendix:catalogues}
Table~\ref{table:eRASSc3catalogue} provides a description of the model for the main and hard catalogues . In both catalogues, each source is assigned a unique source identifier, referred to as {\tt UID} or {\tt DETUID}. Cross-associations between the main and hard sources are recorded in additional columns (e.g. {\tt UID\_Hard}, {\tt UID\_Main}). In a similar fashion, associations with other X-ray catalogues are stored in columns \texttt{UID\_DR1}, \texttt{UID\_5XMM}, \texttt{UID\_2RXS}, and \texttt{UID\_CSC}  (see Sect.~\ref{sec:comparison_Xray_catalogues} for details).

Source properties are generally computed in multiple energy bands, indicated by suffixes in the column names. Band definitions are provided in Tables~\ref{table:energy_bands_main} (from single band detection, 1B) and \ref{table:energy_bands_Hard} (from the three-band detection, 3B). In the 1B detection, only band $1$ ($0.2-2.3$~keV) is involved, and the all-band summary value (band $0$) is identical to that of band $1$. In the 3B detection, the band index $1$, $2$, and $3$ indicate $0.2-0.6$, $0.6-2.3$, and $2.3-5.0$~keV, and band $0$ indicates all-band summary values. Several parameters (e.g. {\tt DET\_LIKE}, {\tt ML\_RATE}, {\tt ML\_CTS}, {\tt ML\_FLUX}, {\tt EXT}, {\tt EXT\_LIKE}) are derived from PSF fitting using the task {\tt ermldet}. Forced PSF fitting ({\tt ermldet}) and aperture photometry ({\tt apetool}) are performed at fixed source positions for bands $P1-P9$. The soft band ($S$; $0.5-2.0$~keV) is obtained by combining bands $P1$ ($0.5-1.0$~keV) and P2 ($1.0-2.0$~keV).
Unlike for the main and hard X-ray catalogues, the column's description for the six catalogues of counterparts is only presented in the data model available on the dedicated eROSITA-DE DR2 website\footnote{ \url{https://erosita.mpe.mpg.de/dr2/AllSkySurveyData\_dr2/Catalogues\_dr2/}}.

\begin{table*}[htp]
  \centering
  \scriptsize
  \caption{eRASS:3 main and hard catalogues column description. All errors are provided as $68\%$ confidence intervals (1$\sigma$).}
  \renewcommand{\arraystretch}{1.1}
  \begin{tabular}{p{0.15\textwidth} p{0.09\textwidth} p{0.68\textwidth}}
   \hline
   \hline
Column         &Units & Description \\
   \hline
\texttt{IAUNAME} & & String containing the official IAU name of the source.\\
\texttt{DETUID} & & String unique detection ID.\\
\texttt{SKYTILE} & & Sky tile ID.\\
\texttt{ID\_SRC} & & Source ID in each sky tile. Use \texttt{SKYTILE+ID\_SRC} to identify the corresponding source products.\\
\texttt{UID} & & Integer unique detection ID. It equals \texttt{CatID$\times 10^{11}$+SKYTILE$\times 10^5$+ID\_SRC}, where \texttt{CatID} is 1 for the 1B detected main, and 2 for the 3B detected hard catalogue.\\
\texttt{UID\_Hard} & & Hard catalogue \texttt{UID} of the source with a strong association, or \texttt{-UID} if the association is weak. 0 means no counterpart found in the hard catalogue. {\it Only in the 1B main catalogue}.\\
\texttt{UID\_Main} & & Main \texttt{UID} of the source with a strong association, or \texttt{-UID} if the association is weak. 0 means no counterpart found in the 1B catalogues. {\it Only in the hard catalogue}.\\
\texttt{UID\_DR1} & & eRASS1 Main catalogue \texttt{UID} of the source with a strong association, or \texttt{-UID} if the association is weak. 0 means no counterpart found in the eRASS1 main catalogue. {\it Only in the 1B main catalogue}.\\
\texttt{UID\_DR1Hard} & & eRASS1 hard \texttt{UID} of the source with a strong association, or \texttt{-UID} if the association is weak. 0 means no counterpart found in the eRASS1 hard catalogue. {\it Only in the hard catalogue}.\\
\texttt{ID\_CLUSTER} & & Group ID of simultaneously fitted sources.\\
\texttt{UID\_5XMM} & & XMM-{\it Newton} 5XMM-DR15 catalogue \texttt{SRCID} of the associated source: positive for strong associations, negative for weak ones; 0 indicates no counterpart, and $-1$ denotes sources outside the 5XMM footprint.\\
\texttt{UID\_2RXS} & & 2RXS catalogue \texttt{IND\_2RXS} of the associated source: positive for strong associations, negative for weak ones; 0 indicates no counterpart.\\
\texttt{UID\_CSC} & & {\it Chandra} CSC 2.1 catalogue \texttt{name} of the associated source: positive for strong associations, negative for weak ones; 0 indicates no counterpart, and $-1$ denotes sources outside the {\it Chandra} CSC 2.1 footprint.\\
\texttt{FLAG\_CSC} & & Association flag for CSC 2.1 (IAU-designated sources): 1 (strong), $-2$ (weak), 0 (no association within the CSC footprint), $-1$ (outside the CSC 2.1 footprint).\\
\hline
\texttt{RA}             	&deg	& Right ascension (ICRS), corrected.\\
\texttt{DEC}            	&deg	& Declination (ICRS), corrected.\\
\texttt{RA\_RAW}       	&deg	& Right ascension (ICRS), uncorrected.\\
\texttt{DEC\_RAW}      	&deg	& Declination (ICRS), uncorrected.\\
\texttt{RA\_LOWERR}        &  arcsec   & 1$\sigma$ lower error of RA.\\
\texttt{RA\_UPERR}         &  arcsec   & 1$\sigma$ upper error of RA.\\
\texttt{DEC\_LOWERR}       &  arcsec   & 1$\sigma$ lower error of Dec.\\
\texttt{DEC\_UPERR}        &  arcsec   & 1$\sigma$ upper error of Dec.\\
\texttt{RADEC\_ERR}     	&arcsec	& Combined positional error, raw output from PSF fitting.\\
\texttt{POS\_ERR} & arcsec & 1$\sigma$ position uncertainty.\\
\texttt{LII} & deg & Galactic longitude.\\
\texttt{BII} & deg & Galactic latitude.\\
\texttt{ELON} & deg & Ecliptic longitude.\\
\texttt{ELAT} & deg & Ecliptic latitude.\\
\hline
\texttt{EXT}            	& arcsec	& Source extent parameter.\\
\texttt{EXT\_ERR}       	& arcsec	& 1$\sigma$ error of \texttt{EXT}.\\
\texttt{EXT\_LOWERR}      & arcsec      & 1$\sigma$ lower error of \texttt{EXT}.\\
\texttt{EXT\_UPERR}       & arcsec      & 1$\sigma$ upper error of \texttt{EXT}.\\
\texttt{EXT\_LIKE}      	&	& Extent likelihood.\\
\texttt{DET\_LIKE\_{\sl\scriptsize n}}  	&	& Detection likelihood. 1B: $n=0$,~$P[1-9]$; 3B: $n=0,\,1,\,2,\,3$.\\
\texttt{ML\_CTS\_{\sl\scriptsize n}}     	&cts	& Source net counts. 1B: $n=1$,~$P[1-9]$; 3B: $n=0,\,1,\,2,\,3$.\\
\texttt{ML\_CTS\_ERR\_{\sl\scriptsize n}} 	&cts	& 1$\sigma$ combined counts error. 1B: $n=1$,~$P[1-9]$; 3B: $n=0,\,1,\,2,\,3$.\\
\texttt{ML\_CTS\_LOWERR\_{\sl\scriptsize n}} 	&cts	& 1$\sigma$ lower counts error. 1B: $n=1$,~$P[1-9]$; 3B: $n=1,\,2,\,3$.\\
\texttt{ML\_CTS\_UPERR\_{\sl\scriptsize n}} 	&cts	& 1$\sigma$ upper counts error. 1B: $n=1$,~$P[1-9]$; 3B: $n=1,\,2,\,3$.\\
\texttt{ML\_RATE\_{\sl\scriptsize n}}    	&cts/s	& Source count rate. 1B: $n=1$,~$P[1-9]$; 3B: $n=0,\,1,\,2,\,3$.\\
\texttt{ML\_RATE\_ERR\_{\sl\scriptsize n}}	&cts/s	& 1$\sigma$ combined count rate error. 1B: $n=1$,~$P[1-9]$; 3B: $n=0,\,1,\,2,\,3$.\\
\texttt{ML\_RATE\_LOWERR\_{\sl\scriptsize n}}	&cts/s	& 1$\sigma$ lower count rate error. 1B: $n=1$,~$P[1-9]$; 3B: $n=1,\,2,\,3$.\\
\texttt{ML\_RATE\_UPERR\_{\sl\scriptsize n}}	&cts/s	& 1$\sigma$ upper count rate error. 1B: $n=1$,~$P[1-9]$; 3B: $n=1,\,2,\,3$.\\
\texttt{ML\_FLUX\_{\sl\scriptsize n}}    	&erg cm$^{-2}$ s$^{-1}$	& Source flux. 1B: $n=1$,~$P[1-9]$; 3B: $n=0,\,1,\,2,\,3$.\\
\texttt{ML\_FLUX\_ERR\_{\sl\scriptsize n}}	&erg cm$^{-2}$ s$^{-1}$	& 1$\sigma$ combined flux error. 1B: $n=1$,~$P[1-9]$; 3B: $n=0,\,1,\,2,\,3$.\\
\texttt{ML\_FLUX\_LOWERR\_{\sl\scriptsize n}}	&erg cm$^{-2}$ s$^{-1}$	& 1$\sigma$ lower flux error. 1B: $n=1$,~$P[1-9]$; 3B: $n=1,\,2,\,3$.\\
\texttt{ML\_FLUX\_UPERR\_{\sl\scriptsize n}}	&erg cm$^{-2}$ s$^{-1}$	& 1$\sigma$ upper flux error. 1B: $n=1$,~$P[1-9]$; 3B: $n=1,\,2,\,3$.\\
\texttt{ML\_BKG\_{\sl\scriptsize n}}     	&cts/arcmin$^2$	& Background at the source position. 1B: $n=1$,~$P[1-9]$; 3B: $n=0,\,1,\,2,\,3$.\\
\texttt{ML\_EXP\_{\sl\scriptsize n}}     	&s	& Vignetted exposure time at the source position. 1B: $n=1$,~$P[1-9]$; 3B: $n=1,\,2,\,3$.\\
\texttt{ML\_EEF\_{\sl\scriptsize n}}     	    &   & Enclosed energy fraction. 1B: $n=1$,~$P[1-9]$; 3B: $n=1,\,2,\,3$.\\
\texttt{APE\_CTS\_{\sl\scriptsize n}}    	&cts	& Total counts extracted within the aperture. 1B: $n=1$,~$P[1-9]$,~$S$; 3B: $n=1,\,2,\,3$.\\
\texttt{APE\_BKG\_{\sl\scriptsize n}}    	&cts	& Background counts extracted within the aperture, excluding nearby sources using the source map. 1B: $n=1$,~$P[1-9]$,~$S$; 3B: $n=1,\,2,\,3$.\\
\texttt{APE\_EXP\_{\sl\scriptsize n}}    	&s	& Exposure map value at the given position. 1B: $n=1$,~$P[1-9]$,~$S$; 3B: $n=1,\,2,\,3$.\\
\texttt{APE\_RADIUS\_{\sl\scriptsize n}} 	&pixels	& Extraction radius in pixels ($4\arcsec$). 1B: $n=1$,~$P[1-9]$; 3B: $n=1,\,2,\,3$.\\
\texttt{APE\_POIS\_{\sl\scriptsize n}}   	&	& Poisson probability that the extracted counts (\texttt{APE\_CTS}) are a background fluctuation. 1B:~$n=1$,~$P[1-9]$,~$S$; 3B: $n=1,\,2,\,3$.\\
\texttt{HR\_{\sl\scriptsize nm}} & & Hardness ratios 1B:~$n=1$,~$P[1-9]$,~$S$. {\it Only in the 1B Main catalogue}. \\
\texttt{HR\_{\sl\scriptsize nm}\_ERR} & & Error on the hardness ratios 1B:~$n=1$,~$P[1-9]$,~$S$. {\it Only in the 1B main catalogue}. \\
\hline
\texttt{FLAG\_SP\_SNR} & & Source might lie within an overdense region near a supernova remnant.\\
\texttt{FLAG\_SP\_BPS} & & Source might lie within an overdense region near a bright point source.\\
\texttt{FLAG\_SP\_SCL} &  & Source might lie within an overdense region near a stellar cluster.\\
\texttt{FLAG\_SP\_LGA} & & Source might lie within an overdense region near a local large galaxy.\\
\texttt{FLAG\_SP\_GC\_CONS} & &  Source might lie within an overdense region near a galaxy cluster.\\
\texttt{FLAG\_NO\_RADEC\_ERR} & &  Source contained no \texttt{RA\_DEC\_ERR} in the pre-processed version of the catalogue.\\
\texttt{FLAG\_NO\_EXT\_ERR} & & Source contained no \texttt{EXT\_ERR} in the pre-processed version of the catalogue.\\
\texttt{FLAG\_NO\_CTS\_ERR} & &  Source contained no \texttt{ML\_CTS\_0\_ERR} in the pre-processed version of the catalogue.\\
\texttt{FLAG\_OPT} & & Source is located within a distance of $15\arcsec$ to an optically bright source.\\
\hline
  \end{tabular}
  \label{table:eRASSc3catalogue}
\end{table*}

\begin{table}
\small
\caption{Dictionary of energy band suffixes in the eRASS:3 main catalogue.}
\centering
\renewcommand{\arraystretch}{1.2}
\begin{tabular}{c c}
\hline
\hline
Band       & Energy range  \\  
\hline	
0, 1 & $0.2-2.3$ keV \\
P1 & $0.2-0.5$ keV \\
P2 & $0.5-1.0$ keV \\
P3 & $1.0-2.0$ keV \\
P4 & $2.0-5.0$ keV \\
P5 & $5.0-8.0$ keV \\
P6 & $4.0-10.0$ keV \\
P7 & $5.1-6.1$ keV \\
P8 & $6.7-7.1$ keV \\
P9 & $7.2-8.2$ keV \\
S & $0.5-2.0$ keV\\
\hline
\end{tabular}
\label{table:energy_bands_main}
\end{table}

\begin{table}
\small
\caption{Dictionary of energy band suffixes in the eRASS:3 hard catalogue.}
\centering
\renewcommand{\arraystretch}{1.2}
\begin{tabular}{c c}
\hline
\hline
Band       & Energy range  \\  
\hline	
0  & $0.2-5.0$ keV \\
1  & $0.2-0.6$ keV \\
2  & $0.6-2.3$ keV \\
3  & $2.3-5.0$ keV \\
\hline
\end{tabular}
\label{table:energy_bands_Hard}
\end{table}

\end{appendix}

\end{document}